\documentclass[12pt,epsfig]{article}
\usepackage{amssymb}
\usepackage{graphicx}

\begin{document}

\newcommand \be  {\begin{equation}}
\newcommand \bea {\begin{eqnarray} \nonumber }
\newcommand \ee  {\end{equation}}
\newcommand \eea {\end{eqnarray}}

\title{Level curvature distribution: from bulk to the soft edge of random Hermitian matrices.\footnote{The text is based on
the presentation at the 5th Workshop on Quantum Chaos and Localization Phenomena, May 20-22, 2011, Warsaw, Poland }}

\vskip 0.2cm
\author{Yan V. Fyodorov$^1$ \\
\noindent\small{$^1$ School of Mathematical Sciences,
University of Nottingham, Nottingham NG72RD, England}}

\maketitle

\begin{abstract}
Level curvature is a measure of sensitivity of  energy levels of a disordered/chaotic system to perturbations.
In the bulk of the spectrum  Random Matrix Theory predicts the probability distributions of level curvatures
to be given by Zakrzewski-Delande expressions [F. von Oppen {\it Phys. Rev. Lett.} {\bf 73} 798 (1994) $\&$ {\it Phys.Rev. E} {\bf 51} 2647 (1995);
 Y.V. Fyodorov and H.-J. Sommers {\it Z.Phys.B} {\bf 99} 123 (1995)]. Motivated by growing interest in statistics of extreme
(maximal or minimal) eigenvalues of disordered systems of various nature, it is natural to ask about the associated level curvatures.
I show how calculating the distribution  for the curvatures of extreme eigenvalues in GUE ensemble
can be reduced to studying asymptotic behaviour of orthogonal polynomials appearing in the recent work C. Nadal and S. N.  Majumdar {\it J. Stat. Mech.} {\bf 2011} P04001 (2011). The corresponding asymptotic analysis being yet outstanding, I instead will discuss solution
 of a related, but somewhat simpler problem of calculating the level curvature distribution averaged over all the levels
in a spectral window close to the edge of the semicircle. The method is based on asymptotic analysis of kernels
 associated with Hermite polynomials and their Cauchy transforms, and is straightforwardly extendable to any
rotationally-invariant ensemble of random matrices.

\end{abstract}

\section{Introduction}

Let $H_N$ stands for $N\times N$ random Hermitian matrix belonging to a certain invariant ensemble which is
 characterized by the joint probability density of $N$ real eigenvalues $\lambda_i, \,,\ i=1,\ldots, N$ of the form
\be\label{Int1}
{\cal P}_N(\lambda_{1},\lambda_2,\ldots,\lambda_N)=\frac{1}{Z_{N}}\,e^{-\frac{N}{2}\sum_{n=1}^NV(\lambda_n)}
\prod_{n<l}^N\,(\lambda_n-\lambda_l)^2\,.
\ee
 in terms of the potential $V(\lambda)$, with $Z_{N}$ being the appropriate normalization constant. In particular, the simplest choice $V(\lambda)=\lambda^2$ corresponds to the so-called Gaussian Unitary Ensemble (GUE) whose mean eigenvalue density is given in the limit $N\to \infty$ by the  Wigner semicircle law $\rho(\mu)=\frac{1}{N}\left\langle\sum_{n=1}^N \delta(\mu-\lambda_n)\right\rangle_{GUE}=\frac{1}{2\pi}\sqrt{4-\mu^2}, \quad |\lambda|<2$.
 This law shows, in particular, that typically the minimal $\lambda_{min}$ and the maximal $\lambda_{max}$ eigenvalues approach $\pm 2$, respectively. For large but finite $N$
  one of the most important chracteristics of GUE spectrum
 appears to be  the Tracy-Widom (TW) law for the  distribution of this {\it extreme} eigenvalues given by \cite{TW}
\be\label{Int2}
{\cal P}(\lambda_{max})=F_2\left(\frac{\lambda_{max}-2}{N^{2/3}}\right), \quad  F_2(x)=\frac{d}{dx}\exp\left[-\int_x^{\infty}(z-x)q^2(z)\,dz\right]
\ee
where $q(z)$ satisfies the Painleve II equation:
\be\label{Int3}
 q''(z)=2q^3(z)+zq(z), \quad q(z\to \infty)\sim Ai(z)\sim \frac{1}{2\sqrt{\pi}z^{1/4}}\, e^{-\frac{2}{3}z^{3/2}}
\ee
Similar distributions are also known for other symmetry classes of random matrices. TW distributions are highly universal, retain their validity not only for
 invariant ensembles with a quite general potential $V(\lambda)$, but also for a very broad class of Hermitian random matrices with independent entries \cite{Soshnikov}, and emerge in several disordered/chaotic physical  systems. E.g they are relevant for describing directed polymers in disordered media \cite{Johansson}, spectral gap fluctuations in disordered metal grains in contact to a bulk superconductor \cite{Vavilov}, fluctuations of output power in coupled fiber lasers \cite{Fridman}, etc.

Consider now a general perturbation $H_N+\gamma W$, where $\gamma$ is the control parameter, and $W$ is a fixed given matrix.
One can pose a natural question of characterizing sensitivity of the minimal/maximal eigenvalue $\lambda_{min}(\gamma)$ to a generic perturbation by
considering the standard perturbation theory: $\lambda_{min}(\gamma)=\lambda_{min}+\gamma\, V+\gamma^2\, C+\ldots$, where
\be\label{1}
V=\langle m|W|m\rangle , \quad C=\sum_{n\ne m}^N\frac{\langle m|W|n\rangle\langle n|W|m\rangle}{\lambda_{min}-\lambda_n}\,.
\ee
Here $|n\rangle,\lambda_n$ for $ n=1,\ldots, N$
 is the set of of eigenvectors/eigenvalues of $H_N$, that is $H|n\rangle=\lambda_n|n\rangle$, and $|m\rangle $ stands for the eigenvector corresponding to the unperturbed minimal eigenvalue $\lambda_{min}$.
The coefficients $V$ and $C$  in (\ref{1}) are frequently called in the physical literature the "level velocity" and the "level curvature", respectively. This terminology is inherited from the use of eigenvalues of random matrices as a model of highly excited energy levels of disordered/chaotic quantum systems, see e.g. \cite{QChaos}. As is well-known, the components of eigenvectors of invariant random matrices are statistically independent from the eigenvalues, and also in the large-$N$ limit behave essentially as independent, identically distributed
  Gaussian variables with variance $1/N$. This makes calculating the distribution of $V$ within the Random Matrix Theory (RMT) context
  a straightforward task.  At the same time finding statistics of the level curvature $C$ is a much less trivial problem.
Various aspects of the level curvatures for eigenvalues {\it in the bulk of the spectrum} of random matrix ensembles , as well as for disordered and chaotic systems attracted quite a considerable interest in mid-'90s. The corresponding curvature distributions were conjectured by Zakrzewski-Delande\cite{ZD} on the basis of numerical simulations, and derived in the limit $N\gg 1$ for Gaussian ensembles in
 \cite{FvO} and independently by a different method in \cite{FS} (see yet another technique in  \cite{EF}). One expects the results to
 be universal, that is to hold for a broad class of random matrices sharing the same global symmetries.

To this end it seems natural to pose questions related to sensitivity of extreme eigenvalues to perturbations. Apart from a generic interest,
 from a somewhat different angle the above expressions characterize sensitivity of the ground state of
the so-called spherical spin-glass model\cite{spherical} to perturbations in random interactions. In addition, one can show
also that in the framework of the same spin-glass-type spherical model the so-called nonlinear susceptibility
of the ground state with respect to external magnetic field can be reduced to a similar, but a more complicated expression.
Those observations provide an additional motivation to try to develop regular tools for statistical characterization of level curvatures for extreme eigenvalues. Here I report some preliminary steps in this direction.

The presentation below will have the following structure. I will start with recapitulating a simpler problem of the level curvature distribution,  with the curvatures being sampled over all levels in a small spectral window around some point in the spectrum.
Departing from the specific  methods used in earlier papers \cite{FvO}, \cite{FS} for Gaussian RMT ensembles, I show that for any rotationally invariant ensemble of Hermitian random matrices the problem can be reduced to the asymptotic analysis of the orthogonal polynomials, and their so-called {\it Cauchy transforms}. After that I will go back to the problem of sensitivity of extreme eigenvalues and show how the corresponding curvature distribution can be formulated in terms of a special class of orthogonal polynomials studied recently in \cite{NM}. The full asymptotic analysis of the resulting kernel is still outstanding,
I will briefly show instead how the new formulation allows to reproduce known GUE
results in the bulk, and then concentrate on the GUE "soft-edge" scaling regime which was not yet studied before.
 Appendices A,B contain some technical details related to
asymptotic analysis of Hermite polynomials.

\section{Level curvature distribution for invariant ensembles: general consideration}

\subsection{Spectrally-averaged curvature distribution in terms of orthogonal polynomials}
The probability density of level curvatures
{\it averaged over all the eigenvalues around the point $\mu$ in the spectrum} is defined as
\be\label{defmeancurv}
{\cal P}(c,\mu)=\frac{1}{N\rho(\mu)}\left\langle\sum_{m=1}^N\delta(c-C_m/C_{typ}) \delta(\mu-\lambda_m)\right\rangle_{H_N}
\ee
where $\langle ...\rangle_H$ stands for the ensemble average, $\rho(\mu)=\frac{1}{N}\left\langle\sum_{n=1}^N \delta(\mu-\lambda_n)\right\rangle_{H_N} $ stands for the mean eigenvalue density around the point $\mu$, the curvature related to the $m$-th eigenvalue being defined as
\be\label{levcurv}
C_m=\sum_{n\ne m}^N\frac{\langle m|W|n\rangle \langle n|W|m\rangle }{\lambda_m-\lambda_n}
\ee
and $C_{typ}$ stands for the typical curvature scale  identified via considering the typical contribution of two neighbouring eigenvalues
\be\label{ctyp}
C_{typ}=\frac{1}{\Delta}\overline{\langle m|W|n\rangle \langle n|W|m\rangle }=\pi \rho(\mu)\,y_{typ}, \quad y_{typ}=\frac{1}{N}\mbox{Tr}W^2
\ee
with the parameter $\Delta=1/(N\pi \rho(\mu))$ defining the mean spacing between the neighbouring eigenvalues.
The bar stands for the  average over eigenvectors, and we assumed above that both $|n\rangle$ and $|m\rangle$ are independent $N-$component gaussian
complex-valued vectors with mean unit length (see below). In what follows we also assume that a {\it generic} perturbations
is such that the variable $y_{typ}=O(1)$ for $N\to \infty$. More precisely,  we consider only full-rank perturbation $W$ such that all its $N$ eigenvalues are of the order of unity.

Let us note for the future use that from the definitions (\ref{defmeancurv},\ref{levcurv})  the mean curvature for levels around a point $\mu$ in the spectrum is given by
\be\label{curav}
\left\langle C_m\right\rangle_H=y_{typ} \frac{1}{N^2\rho(\mu)}\int_{-\infty}^{\infty} {\cal R}_2(\mu,\lambda)\frac{d\lambda}{\mu-\lambda}
\ee
where we introduced the standard eigenvalue two-point cluster function
\be\label{defr2}
 {\cal R}_2(\mu,\lambda)=N(N-1)\int_{-\infty}^{\infty} {\cal P}_N\left(\lambda_1=\mu,\lambda_2=\lambda,\lambda_3,\ldots,\lambda_N\right)\,d\lambda_3\ldots \lambda_N
\ee
Note that the eigenvalues are strongly correlated only over the distance comparable with typical level spacing $\Delta$ which is negligible in comparison with the total length of the spectral support (assumed to be for simplicity a single interval).  It is then easy to see that the leading order result is obtained by neglecting the correlations and using $ {\cal R}_2(\mu,\lambda)\approx N^2\rho(\mu)\rho(\lambda)$ in (\ref{curav}). In particular, for GUE  the mean curvature in the limit $N\to \infty$ is simply given by
\be\label{curavGUE}
\left\langle C_m\right\rangle_H=y_{typ}\left\{\begin{array}{c} \frac{\mu}{2}, \quad |\mu|<2\\ \frac{\mu-\sqrt{\mu^2-4}}{2}, \quad |\mu|>2 \end{array}\right.
\ee
where we have assumed that $\mu$ is fixed when $N\to \infty$, and used the known formula for the averaged resolvent of the GUE matrix.
We also included for further reference second line corresponding formally to the situation when the observation window is chosen to be outside the support.

After this digression let us proceed to calculating the Fourier transform ${\cal K}(\omega)= \int {\cal P}(c,\mu)e^{-i\omega c}\, dc$
of the probability density for normalized curvatures $c=C/C_{typ}$. Introducing for notational shorthand  the vector $|W_m\rangle =W|m\rangle$, we have after straightforward manipulations:
\be\label{charfun1}
{\cal K}(\omega)=\frac{1}{N\rho(\mu)}\left\langle\sum_{m=1}^N \delta(\mu-\lambda_m)\prod_{n(\ne m)}\exp\left\{
-i\frac{\omega}{C_{typ}}\frac{\langle n|W_m\rangle\langle W_m|n\rangle}{\mu-\lambda_n}\right\}\right\rangle_{H_N}
\ee

We will perform the ensemble average in steps, and start with averaging over the eigenvectors $|n\rangle$ with $n=1,2,\ldots,m-1,m+1,\ldots, N$.
In doing this we assume that in the large$-N$ limit different eigenvectors are effectively statistically independent for different $n$,
 and moreover the components $n_i=\langle i|n\rangle$ of a given eigenvector in any basis $|i\rangle$  can be treated in the same limit
 simply as i.i.d. complex Gaussian-distributed numbers with variance $1/N$. Actually, one can relatively easily verify
 that if one takes any finite subset of $l$ eigenvectors such that $l$ is fixed when $ N\to \infty$  then the required properties follow.
 The latter statement is rigorously proved in the mathematical literature, see e.g.\cite{Jiang}.
  Therefore, our method implicitly assumes that the number $l$ of terms which effectively contribute
 to the curvature defined in (\ref{levcurv}) is much smaller than $N$. This is very plausible in view of the denominators growing roughly linearly
 with $n$, but strictly speaking remains a conjecture. Using such an assumption we can easily perform the eigenvector average for a given $|n\rangle$, which simply amounts to using the Gaussian identity:
\[
\frac{1}{\cal N}\int e^{-N\langle n|n\rangle-i\alpha_n\,\langle n|W_m\rangle\langle W_m|n\rangle} \,d^2|n\rangle\,= \frac{1}{\det({\bf 1}+i\frac{\alpha_n}{N}|W_m\rangle\langle W_m|)}
 \]
\be
 =\frac{1}{1+i\frac{\alpha_n}{N}\langle W_m|W_m\rangle}, \quad \alpha_n=\frac{\omega}{C_{typ}}\frac{1}{\mu-\lambda_n}
\ee
This immediately allows us to write
\be\label{carfun2}
{\cal K}(\omega)= \frac{1}{N\rho(\mu)}\overline{\left\langle\sum_{m=1}^N \delta(\mu-\lambda_m)\prod_{n\ne m}^N\frac{\mu-\lambda_{n}}
{\mu-\lambda_{n}+i\omega \frac{y_m}{N} }\right\rangle_{\lambda_n}}
\ee
where the brackets now stand for the averaging over the joint probability density (\ref{Int1}), and we have introduced (random) variable $y_m=\frac{\langle W_m|W_m\rangle}{ C_{typ}}\equiv \frac{\langle m|W^2|m\rangle}{ C_{typ}}$ (and exploited the Hermiticity $W=W^{\dagger}$), with the bar standing for the remaining average over that variable.

Let us briefly discuss how one performs the latter average by calculating the distribution of the variable $y_m$.
Exploiting the mentioned Gaussianity of the individual eigenvector we immediately find for the corresponding characteristic function:
 \be\label{delta}
 \overline{\exp\left\{i\frac{q}{C_{typ}}\langle m|W^2|m\rangle\right\}}=\frac{1}{\det({\bf 1}-i\frac{q}{NC_{typ}}W^2)}\approx
 e^{i\frac{q}{NC_{typ}}\mbox{\small Tr}W^2-\frac{q^2}{2N^2C_{typ}^2}\mbox{\small Tr}W^4+\ldots}
 \ee
As for a "generic" perturbation we must have $\mbox{ Tr} W^{2p}=O(N)$ for all integer $p\ge 1$ one can for large $N$ effectively retain only the first term in the exponential in (\ref{delta}). We conclude that
 the distribution of $y_m$ is effectively $\delta-$functional in the limit $N\gg 1$, so one can replace $y_m$
 with its average value: $y_m\to \overline{y_m}=\frac{1}{NC_{typ}}\mbox{ Tr} W^{2}=\frac{1}{\pi\rho(\mu)}\equiv \Delta N$. This gives
 \be\label{carfun3}
{\cal K}(\omega)= \frac{1}{\rho(\mu)}\left\langle \delta(\mu-\lambda_N)\prod_{n=1}^{N-1}\frac{\mu-\lambda_{n}}
{\mu-\lambda_{n}+i\omega \Delta }\right\rangle_{\lambda_1,\ldots,\lambda_N}
\ee
where we have used that by permutation symmetry of the joint probability density (\ref{Int1}) all the $N$ terms in (\ref{carfun2}) produce identical contribution upon averaging,
so it is sufficient to consider averaging of a single term with $m=N$. To perform the remaining averaging over eigenvalues it is convenient to introduce the ensemble of $(N-1)\times (N-1)$ Hermitian matrices $H_{N-1}$ whose joint probability density of eigenvalues is given by (cf.(\ref{Int1}) )
\be\label{Int2a}
{\cal P}_{N-1}(\lambda_{1},\lambda_2,\ldots,\lambda_{N-1})=\frac{1}{Z_{N-1}}\,e^{-\frac{N}{2}\sum_{n=1}^{N-1}V(\lambda_n)}
\prod_{n<l}^{N-1}\,(\lambda_n-\lambda_l)^2\,.
\ee
This allows us to rewrite (\ref{Int1}) as
\be\label{Int3a}
{\cal P}_{N}(\lambda_{1},\lambda_2,\ldots,\lambda_{N})=\frac{Z_{N-1}}{Z_N}e^{-NV(\lambda_N)}\prod_{n=1}^{N-1}\left(\lambda_N-\lambda_{n}\right)^2{\cal P}_{N-1}(\lambda_{1},\lambda_2,\ldots,\lambda_{N-1})
\ee
 Introducing the characteristic polynomial $\det{(\mu-H_{N-1})}$ one can notice that the integration over $\lambda_1,\ldots, \lambda_{N-1}$ in (\ref{carfun3}) amounts to evaluating the following object
\be\label{6}
\left\langle\frac{\left[\det{(\mu-H_{N-1})}\right]^3}{\det{(\epsilon-H_{N-1})}}\right\rangle_{H_{N-1}}\equiv \frac{2\pi i}{[c_{N-1}]^2}{\cal F}(\mu,\epsilon)
\ee
where we have denoted $\epsilon=\mu+i\omega\,\Delta, \quad \Delta=\frac{1}{\pi\rho(\mu)N}$ and the constant coefficients $[c_{N-1}]^2$ will be defined below. In terms of the above function the Fourier-transform of the curvature distribution is simply given by
\be\label{curvdis}
\int {\cal P}(c,\mu)e^{-i\omega c} dc =\frac{1}{\rho(\mu)}\frac{Z_{N-1}}{Z_{N}}e^{-\frac{N}{2}\mu^2}\frac{2\pi i}{[c_{N-1}]^2}\,{\cal F}(\mu,\mu+i\omega \Delta)
 \ee
The correlation functions of product/ratios of characteristic polynomials of random matrices, like that featuring in (\ref{6}),
 were considered in full generality in \cite{FStr,StrF}. There it was found how to express these objects in terms
of orthogonal polynomials generated by the potential $V(x)$. Namely, introducing a measure on the real line as $d\mu(x)=e^{-NV(x)}\,dx$, one can define the unique set $\pi_k(x)$ of associated {\it monic} orthogonal polynomials satisfying
\be\label{7}
\int \pi_k(x)\pi_j(x)d\mu (x)=\left[c_k\right]^2\delta_{jk}
\ee
As is well-known (see e.g. \cite{Deift,mylec} and references therein) the product of the coefficients $[c_k]^2$ gives the normalization constant
in (\ref{Int1}):  $Z_{N-1}=(N-1)!\prod_{k=0}^{N-2}\left[c_k \right]^2$.
Further, define the so called {\it Cauchy transforms}
\be\label{8}
h_k(\epsilon)=\frac{1}{2\pi i}\int \frac{\pi_k(x)}{x-\epsilon}\, d\mu(x), \quad Im{\epsilon}\ne 0
\ee
In particular, in \cite{StrF} an expression
was derived relating general averages like those featuring in (\ref{6}) to the determinants of $2\times 2$ matrices whose entries are certain
 bi-linear combinations of the polynomials and their derivatives (the so-called "kernels"). After specifying the general formulae for our particular case the correlation function defined in (\ref{curvdis}) takes the following explicit form
\be\label{6bb} {\cal F}(\mu,\epsilon)=
F_1(\mu,\epsilon)\left[W_1(\mu,\mu)-\frac{1}{2}(\epsilon-\mu)W_2(\mu,\mu)\right]+(\epsilon-\mu)F_2(\mu,\epsilon)W_1(\mu,\mu) \,,
\ee
in terms of the following kernels:
\be\label{F1}
F_1(\mu,\epsilon)= h_{N}(\epsilon)\pi_{N-1}(\mu) - h_{N-1}(\epsilon)\pi_{N}(\mu)
\ee
\be\label{F2}
 F_2(\mu,\epsilon)= h_{N}(\epsilon)\pi'_{N-1}(\mu) - h_{N-1}(\epsilon)\pi'_{N}(\mu)
\ee
and
\bea\label{9b}
 W_1(\mu,\mu)=\pi'_{N}(\mu)\pi_{N-1}(\mu)- \pi_{N}(\mu)\pi'_{N-1}(\mu)&&\\ \quad W_2(\mu,\mu)=\pi''_{N}(\mu)\pi_{N-1}(\mu)- \pi_{N}(\mu)\pi''_{N-1}(\mu)\equiv \frac{d}{d\mu} W_1(\mu,\mu)&&
\eea

So formally the problem amounts to finding asymptotic approximations for the orthogonal polynomials and Cauchy transforms for a given potential  in the specified spectral regime. Various techniques are available for performing such an analysis, the Riemann-Hilbert approach (see e.g. \cite{Deift}) being the most powerful, especially for proving the universality of the required asymptotics for a broad class of rotationally-invariant ensembles.
We will not pursue this line here but
rather show later on how the above formulae reproduce the known Zakrzewski-Delande expressions in the bulk of the spectrum
 $|\mu|<2$ of the Gaussian Unitary ensemble in the large-$N$ limit, and then study the soft-edge case. Before doing that we however
 come back to addressing the original problem of the curvature distribution for extreme eigenvalues, and formulate it in terms of asymptotics of a special class of orthogonal polynomials.

\subsection{Curvature distribution for extreme eigenvalues: orthogonal polynomial formulation}
Consider again the perturbation of the extreme eigenvalue  ($\lambda_{min}$ for definiteness) with the curvature given by (\ref{1}) and normalized to the typical curvature at the soft edge  $\tilde{C}_{typ}= N^{-1/3}y_{typ}$ (see Sec. 3.2 below):
 \be \label{curmindef}
 {\cal P}_m(c)=\left\langle\sum_{m=1}^N\delta\left(c-\frac{1}{\tilde{C}_{typ}}\sum_{n\ne m}^N\frac{\langle m|W|n\rangle \langle n|W|m\rangle }{\lambda_m-\lambda_n}\right)\prod_{n\ne m}^N\chi_{\small \left(\lambda_n>\lambda_{m}\right)} \right\rangle_{H_N}
 \ee
 \be
 =N\left\langle\delta\left(c-\frac{1}{\tilde{C}_{typ}}\sum_{n=2}^N\frac{\langle 1|W|n\rangle \langle n|W|1\rangle }{\lambda_1-\lambda_n}\right)\prod_{n=2}^N\chi_{\small \left(\lambda_n>\lambda_{1}\right)} \right\rangle_{H_N}
 \ee
 where averaging goes over the joint probability density (\ref{Int1}), and
 we have introduced the indicator function: $\chi_{(A)}=1$ if  $A$ is true and zero otherwise, and exploited the permutation symmetry of (\ref{Int1}). Introducing the corresponding Fourier-transform (known as the characteristic function)  ${\cal K}_{m}(\omega)=\left\langle e^{i\omega c}
 \right\rangle $ and  averaging it over the Gaussian eigenvectors $|n\rangle$, with $n=2,3, \ldots, N$, and then over the remaining eigenvector $|1\rangle$ corresponding to $\lambda_{min}$ yields, in full analogy to (\ref{carfun3})
 \be\label{2min}
{\cal K}_{m}(\omega)= \left\langle\prod_{n=2}^N\frac{\lambda_{1}-\lambda_{n}}
{\lambda_{1}-\lambda_{n}+i\frac{\omega}{N^{2/3}}} \prod_{n=2}^N\chi_{\small \left(\lambda_n>\lambda_{1}\right)}\right\rangle_{H_N},
\ee
At the next step we introduce an ensemble of $(N-1)\times (N-1)$  random Hermitian matrix $M_{N-1}$ with the eigenvalues $\lambda_2, \ldots, \lambda_{N}$ and
the (normalized) measure
\be\label{4min}
P_{\lambda_{min}}(M_{N-1})=\frac{1}{Z_{N-1}(\lambda_{1})}\, e^{-\frac{N}{2}\sum_{n=2}^NV\lambda_n)}
\prod_{n=2}^{N}\chi_{\small \left(\lambda_n>\lambda_{1}\right)}\prod_{n<p}^{N-1}(\lambda_n-\lambda_p)^2
\ee
where $Z_{N-1}(\lambda_{1})$ is the appropriate normalization constant.
This distribution (and its normalisation) depends on $\lambda_1(\equiv \lambda_{min})$ as an external parameter. In terms of such an ensemble
we easily see (cf. (\ref{6}))
\be\label{5min}
{\cal K}_m(\omega)= \frac{1}{Z_N}\int_{-\infty}^{\infty}\,d\lambda_{min}e^{-\frac{N}{2}V(\lambda_{min})}Z_{N-1}(\lambda_{min})
\left\langle\prod_{n=1}^{N-1}\frac{(\lambda_{min}-\lambda_{n})^3}{\lambda_{min}-\lambda_{n}+i \frac{\omega}{N^{2/3}}}\right\rangle_{M_{N-1}}
\ee
where the averaging $\left\langle ... \right \rangle_{M_{N-1}}$ goes over the probability density (\ref{4min}).

We conclude that the analysis of the above expression amounts to studying the orthogonal polynomials generated by the measure $P_{\lambda_{min}}(M_{N-1})$. Indeed, we can introduce the measure $d\mu_{\lambda_{min}}(x)=e^{-\frac{N}{2}V(x)}\chi_{(x>\lambda_{min})}\,dx$,  with $\pi_k(x;\lambda_{min})$ standing for monic orthogonal polynomials with respect to this measure satisfying
\be\label{7b}
\int \pi_k(x;\lambda_{min})\,\pi_j(x;\lambda_{min})\,d\mu_{\lambda_{min}}(x)=\left[c_k(\lambda_{min})\right]^2\delta_{jk}
\ee
The above polynomials depend on $\lambda_{min}$ as external parameter, and in this way the analysis of the level curvature distribution
for the minimal eigenvalue amounts to extracting the large-N asymptotic behaviour
of such polynomials and their Cauchy transforms for $N\to \infty$. Several important steps in such an analysis
for the Gaussian case $V(x)=x^2$ were  reported recently by Nadal and Majumdar in \cite{NM}, but further work is needed to
include the Cauchy transforms into considerration to be able to extract the ensuing curvature distribution in the explicit form. The problem is non-trivial and is currently under investigation \cite{FNM}. Below we return to considering a somewhat simpler case of the spectral-averaged curvature distribution, both at the edge and in the bulk.

\section{Spectral-averaged GUE curvature distribution}
\subsection{Bulk of the GUE spectrum}

Denote $\pi_k(x)\equiv p_k(x)=x^k+\ldots$ the monic orthogonal polynomials  w.r.t. the standard Gaussian measure on the full line  $d\mu_{\lambda_{min}}(x)=e^{-\frac{N}{2}x^2}\,dx$, that is
  \be\label{7a}
\int_{-\infty}^{\infty} p_k(x)p_j(x)e^{-\frac{N}{2}x^2}\,dx=c_k^2\delta_{jk}, \quad c_k^2= \frac{k!}{N^k}\sqrt{\frac{2\pi}{N}}
\ee
 Those are actually the classical Hermite polynomials. Renaming the spectral parameter $\mu\to x$, and $y=\frac{1}{\pi\rho(x)}$ so that $\Delta=\frac{y}{N}$ our goal is to calculate (see (\ref{curvdis})
  \be\label{defmeancurv1}
\int {\cal P}(c,x)e^{-i\omega c}dc =  i\frac{N^{2N-1}}{[(N-1)!]^2}\frac{1}{N\rho(x)} e^{-\frac{N}{2}x^2}\,{\cal F}(x,x+i\frac{\omega}{N}y)
  \ee
Relevant bulk asymptotic expressions for Hermite polynomials, Cauchy transforms, and the kernels involved in the curvature distributions are well-known, but to make the present text self-contined I recover them in the Appendix A directly from the integral representations.
Parametrizing a bulk point of the spectrum $x=2\cos{\phi}\in (-2,2)$ for $\eta,\zeta$ of the order of unity and $N\gg 1$ we have
\be\label{ratio}
 -\frac{2\pi i}{c_{N-1}^2}F_1\left(x+\frac{\eta}{N},x+\frac{\zeta}{N}\right)\to e^{-\frac{x}{2}(\zeta-\eta)+i\pi\rho(x) (\zeta-\eta) s_{\zeta}}
\ee
where we denoted $s_{\zeta}=\mbox{sign} Im{(\zeta)}$.

Taking into account $-\frac{c_{N-1}^2}{2\pi i}\to ie^{-N} $ we conclude that
the required large$-N$ asymptotics for any real $\omega$ and $y>0$ of the order of unity  is given by
\be\label{F1as}
F_1\left(x,x+i\frac{\omega}{N}y\right)\approx i \,e^{-N} e^{-i\omega\frac{x}{2}y-\pi \rho(x)|\omega|y}, \,\,
\ee
whereas in the view of the exact relation $F_2\left(x,x+i\frac{\omega}{N}y\right)=N\frac{\partial}{\partial \eta}|_{\eta=0}F_1\left(x+\frac{\eta}{N},x+\frac{\zeta}{N}\right) $ we have
\be\label{F2as}
F_2\left(x,x+i\frac{\omega}{N}y\right)\approx N\,\left[\frac{x}{2}-is_{\omega}\pi\rho(x)\right]\,F_1\left(x,x+i\frac{\omega}{N}y\right)
\ee
We also have
\be\label{kerker}
W_1(x,x)\approx 2N e^{-N}\pi \rho(x) e^{\frac{N}{2}x^2}, \quad W_2(x,x)\approx N\,x\, W_1(x,x)
\ee
Substituting all this to (\ref{6bb})
with $\mu\equiv x, \, \epsilon\equiv x+i\frac{\omega}{N}y$ we find that
\[
e^{-\frac{N}{2}x^2} {\cal F}\left(x,x+i\frac{\omega}{N}y\right)=
\]

\[
 =e^{-\frac{N}{2}x^2} W_1(x,x)F_1\left(x,x+i\frac{\omega}{N}y\right)
\left\{\left[1-\frac{1}{2}i\frac{\omega}{N}\,y\, N\, x\right]+i\frac{\omega}{N}\,y\,N\left[\frac{x}{2}-is_{\omega}\pi\rho(x)\right] \right\}
\]

\be \label{6bbas}
=2i\,N e^{-2N}\pi \rho(x)\,e^{-i\omega\frac{x}{2}y-\pi \rho(x)|\omega|y}(1+y|\omega|\pi\rho(x))
\ee

Now we substitute all this into (\ref{defmeancurv1}) and use the Stirling approximation $(N-1)!=\sqrt{\frac{2\pi}{N}}N^Ne^{-N}$.  Finally we arrive at
\be\label{bulkcurvdiss}
\int {\cal P}(c,x)e^{-i\omega c}dc =  N \,e^{-i\omega\frac{x}{2}y-\pi \rho(x)|\omega|y}(1+y|\omega|\pi\rho(x))\,.
  \ee
This is indeed exactly the expression leading after the Fourier-transform to the Zakrzewski-Delande curvature distribution for GUE ensemble,
see \cite{ZD,FvO,FS}:
\be\label{DZ}
{\cal P}(c,x)=\frac{2}{\pi}\frac{\kappa^3}{\left[(c-c_0)^2+\kappa^2\right]^2}, \quad \kappa=\pi\rho(x)y=1, \,\, c_0=\frac{x}{2}\,y=\frac{x}{2\pi\rho(x)}
\ee

{\bf Note:} The above calculation can be further shortened if we first notice two useful identities, the first one being
 \be\label{denident}
 N\rho(\mu)=\frac{Z_{N-1}}{Z_{N}}e^{-\frac{N}{2}\mu^2}\left\langle\left[\det{(\mu-M_{N-1})}\right]^2\right\rangle_{M_{N-1}}
 \ee
 and second one $ W_1(\mu,\mu)=\left\langle\left[\det{(\mu-M_{N-1})}\right]^2\right\rangle_{M_{N-1}}$ (see e.g. \cite{mylec}). When
combined together they produce the following relation
\be\label{denident1}
 \frac{1}{N\rho(\mu)}\frac{Z_{N-1}}{Z_{N}}e^{-\frac{N}{2}\mu^2}W_1(\mu,\mu)=1\,,
\ee
which after being substited to (\ref{6bb}) helps to rewrite the Fourier-transformed curvature distribution in the most concise form:
\be\label{curvshort}
\int_{-\infty}^{\infty} {\cal P}(c,\mu)e^{-i\omega c} dc =\tilde{{\cal F}}\left(\mu,\mu+i\frac{\omega}{N}y\right)\,,
 \ee
 where we have defined
 \be\label{Fshort}
  \tilde{{\cal F}}(\mu,\epsilon)=
\tilde{F}_1(\mu,\epsilon)\left[1-\frac{1}{2}(\epsilon-\mu)\frac{W_2(\mu,\mu)}{W_1(\mu,\mu)}\right]+(\epsilon-\mu)\tilde{F}_2(\mu,\epsilon) \,,
 \ee
 and $\tilde{F}_{1,2}(\mu,\epsilon)\equiv \frac{2\pi i}{c_{N-1}^2}{F}_{1,2}(\mu,\epsilon)$.

 In particular, such a form turns out to be more convenient for extending the calculation to the spectral-averaged curvature distrubution at the "edge of spectrum" regime which is to be considered in the next section.

 \subsection{Soft edge of the GUE spectrum}
 Consider  the "soft edge" regime $\mu\equiv x=2+\frac{\zeta}{N^{2/3}}$, where $\zeta\in (-\infty, \infty)$.  The mean density of eigenvalues $\rho(x)$ in this regime is well-known and scales as
 \be
  \rho(x)=\frac{1}{N^{1/3}}\tilde{\rho}(\zeta), \quad \mbox{with} \quad
\tilde{\rho}(\zeta)=Ai'(\zeta)^2-Ai(\zeta)Ai''(\zeta)
\ee
where $Ai(\zeta)$ stands for the Airy function. The corresponding mean level spacing is then $\Delta=\frac{\tilde{y}}{N^{2/3}},$ where $\tilde{y}=\frac{1}{\pi\tilde{\rho}(\zeta)}$.
 Relevant soft edge asymptotics of Hermite polynomials, Cauchy transforms, and the kernels involved were considered, for example,
 in \cite{AF} (see also \cite{BHedge}, \cite{Fedge} and \cite{mylec}).  For completeness, we reproduce them in detail in the Appendix B,
see in particular expressions (\ref{softedgeratio}) for $\frac{W_2(x,x)}{2W_1(x,x)}$
and also equations (\ref{F2soft}), (\ref{F1soft}) for $F_2\left(x,x+\frac{i\omega \tilde{y}}{N^{2/3}}\right)$ and $F_1\left(x,x+\frac{i\omega \tilde{y}}{N^{2/3}}\right)$. Remembering $\epsilon-\mu=i\frac{\omega \tilde{y}}{N^{2/3}}$ we reduce (\ref{curvshort},\ref{Fshort}) after a simple algebra to
 \[
 \int {\cal P}(c,\mu)e^{-i\omega c} dc = \tilde{{\cal F}}\left(2+\frac{\zeta}{N^{2/3}},2+\frac{\zeta+i\omega \tilde{y}}{N^{2/3}}\right)\propto -i\pi e^{-i\,N^{1/3}\omega\, \tilde{y}}
 \]
 \be\label{Fshortsoft}
\times\left\{ s_{\omega}\Phi(\zeta,\omega)+i|\omega|\tilde{y}\left[\Psi(\zeta,\omega)-\frac{1}{2}\Phi(\zeta,\omega)\,\frac{\partial}{\partial \zeta}\ln{\tilde{\rho}(\zeta)}\,\right]\right\}
 \ee
 where
\be\label{PhiPsi}
\Phi(\zeta,\omega)=\alpha(\zeta,\omega)\,Ai'(\zeta)-\alpha'(\zeta,\omega)\,Ai(\zeta),\,\,
\Psi(\zeta,\omega)
\ee
\[
=\alpha(\zeta,\omega)\,Ai''(\zeta)-\alpha'(\zeta,\omega)\,Ai'(\zeta)\,
\quad \mbox{with} \quad\,\,\alpha'(\zeta,\omega)\equiv \frac{\partial}{\partial \zeta}\alpha(\zeta,\omega), \, etc.
\]

The exponential factor $e^{-i\,N^{1/3}\omega\, \tilde{y}}$ simply fixes the constant shift in the curvatures $c=C/C_{typ}$ to be
$c_0=N^{1/3} \tilde{y}=\frac{N^{1/3}}{\pi \rho(\zeta)}$ (remembering the correspondence $y\equiv N^{1/3}\tilde{y}$ this value coincides with the mean bulk value $c_0=\frac{x}{2\pi\rho_{av}}$ in the edge limit $x\to 2$). To get rid of such a shift and of the related extra $\rho(\zeta)$ dependence in typical curvature we redefine "shifted and scaled" curvatures $c_{sc}=\frac{c-c_0}{\tilde{y}}\equiv \pi\rho(\zeta)(c-c_0)$ for eigenvalues
around the point $x=2+\frac{\zeta}{N^{2/3}}$. This will allow us to omit the factor  $e^{-i\,N^{1/3}\omega\, \tilde{y}}$
and  set $\tilde{y}=1$ in the definition (\ref{anyomega}) for $\alpha(\zeta,\omega)$.

Note that by setting in (\ref{curavGUE}) $\mu=2+\frac{\zeta}{N^{2/3}}$ yields
for the averaged curvature $\langle C\rangle/y_{typ}\approx 1-\frac{\sqrt{\zeta}}{N^{1/3}}$. Though formally (\ref{curavGUE}) is not valid for the soft edge scaling, and should be replaced by an accurate formula for the mean resolvent involving second solution of the Airy equation $Bi(\zeta)$,  the above estimate works well for
$\zeta\gg 1$). This gives for the mean value
\be
\langle c_{sc}\rangle=\frac{\langle c\rangle-c_0}{\tilde{y}}\equiv N^{1/3}\left(\frac{\langle C\rangle }{y_{typ}}-1\right)\approx -\sqrt{\zeta}
\ee
where we have used $C_{typ} c_0=y_{typ}$. We will indeed find below that $\langle c_{sc}\rangle\approx -\sqrt{\zeta}$, is the most
probable value of the curvature for $\zeta\gg 1$. In what follows we will set again $c_{sc}\equiv c$ for brevity.

Now we proceed to evaluating the full curvature distribution in the soft edge scaling limit. From (\ref{anyomega}) we have
\be\label{alpha}
s_{\omega}\alpha(\zeta,\omega)=\int_0^{\infty}\,d\tau\, \cos{\left(\tau\zeta+\frac{\tau^3}{3}\right)}\, s_{\omega}e^{-|\omega|\tau}
\ee
\[
+i\left[
\int_0^{\infty}\,d\tau\,\sin{\left(\tau\zeta+\frac{\tau^3}{3}\right)}e^{-|\omega|\tau}+
\int_0^{\infty}\,d\tau\,e^{\tau\zeta-\frac{\tau^3}{3}}e^{i\omega\tau}\right]
\]
Further defining
\be\label{beta}
\beta(c,\zeta)=-i\int_{-\infty}^{\infty}\frac{d\omega}{2\pi}\,e^{i\omega c}\,s_{\omega}\alpha(\zeta,\omega)\,
\ee
and using (\ref{alpha}) we then find
\be \label{betaexplicit}
\beta(c,\zeta)=c\gamma(c,\zeta)-\gamma'(\zeta,c)+\delta(c,\zeta)
\ee
where
\be\label{gamma}
\gamma(c,\zeta)=\frac{1}{\pi}\int_0^{\infty}\, \cos{\left(\tau\zeta+\frac{\tau^3}{3}\right)}\frac{d\tau}{c^2+\tau^2},\quad \gamma'(\zeta,c)\equiv \frac{\partial}{\partial \zeta}\gamma(c,\zeta)
\ee
and
\be\label{delta1}
 \delta(c,\zeta)=\theta(-c)e^{-c\,\zeta+\frac{c^3}{3}}\,.
\ee
This yields the contribution to the level curvature distribution corresponding to the first term in (\ref{Fshortsoft})
\be\label{P1}
{\cal P}^{(I)}(c,\zeta)= \left[Ai(\zeta)\,\beta'(c,\zeta)-Ai'(\zeta)\,\beta(c,\zeta)\right]
\ee
where $\beta'(\zeta,c)\equiv \frac{\partial}{\partial \zeta}\beta(c,\zeta)$.

Note that relations (\ref{Ai}, \ref{Bi}) imply
\be\label{check}
\int_{-\infty}^{\infty}\,dc\, \beta(c,\zeta)=\pi Bi(\zeta) \quad \mbox{and}\quad Ai(\zeta)Bi'(\zeta)- Ai'(\zeta)Bi(\zeta)=\frac{1}{\pi}
\ee
which when used together with (\ref{P1}) ensure that the above piece contains the full normalization: $\int_{-\infty}^{\infty}\,dc\,{\cal P}^{(I)}(c,\zeta)=1$.
Now we further notice
\be\label{betader}
\frac{\partial}{\partial c}\beta(c,\zeta)=\int_{-\infty}^{\infty}\frac{d\omega}{2\pi}\,e^{i\omega c}\,|\omega|\alpha(\zeta,\omega)
\ee
which implies from (\ref{PhiPsi})
\be\label{Phifour}
\int_{-\infty}^{\infty}\frac{d\omega}{2\pi}\,e^{i\omega c}\,|\omega|\Phi(\zeta,\omega)=\frac{\partial}{\partial c}\left[Ai'(\zeta)\beta(c,\zeta)-\beta'(\zeta,c)\,Ai(\zeta)\right]\equiv -\frac{\partial}{\partial c}{\cal P}^{(I)}(c,\zeta),
\ee
 Similarly,
\be\label{Psifour}
\int_{-\infty}^{\infty}\frac{d\omega}{2\pi}\,e^{i\omega c}\,|\omega|\Psi(\zeta,\omega)=\frac{\partial}{\partial c}\left[\beta(\zeta,c)\,Ai''(\zeta)-\,Ai'(\zeta)\beta'(c,\zeta)\right],
\ee
Taken together this gives second contribution to the curvature distribution:
 \be\label{P2}
{\cal P}^{(II)}(c,\zeta)= -\frac{\partial}{\partial c}\left\{\beta(\zeta,c)\,Ai''(\zeta)-\,Ai'(\zeta)\beta'(c,\zeta)+{\cal P}^{(I)}(c,\zeta)\frac{1}{2}\frac{\partial}{\partial \zeta}\ln{\rho(\zeta)}\right\}
\ee
The derivative form ensures $\int_{-\infty}^{\infty}\,dc\,{\cal P}^{(II)}(c,\zeta)=0$, as expected. Note also that
using $Ai''(\zeta)=\zeta \,A_i(\zeta)$ it is easy to check $\rho'(\zeta)=-\left[Ai(\zeta)\right]^2$.

Note that from the definitions (\ref{gamma},\ref{delta1}) we have:
\be\label{differ}
\delta'(c,\zeta)=-c\,\delta(c,\zeta), \quad \gamma''(c,\zeta)=c^2\gamma(c,\zeta)-Ai(\zeta)
\ee
which yields
\be\label{betaprim}
\beta'(c,\zeta)=-c\,\beta(c,\zeta)+Ai(\zeta)\,.
\ee
Substituting this to (\ref{P1}) we get
\be\label{P1simp}
{\cal P}^{(I)}(c,\zeta)=-\beta(c,\zeta)\left[c\,Ai(\zeta)+Ai'(\zeta)\right]+Ai^2(\zeta)
\ee
which implies:
\be\label{P2partB}
\frac{\partial}{\partial c}{\cal P}^{(I)}(c,\zeta)=-\frac{\partial}{\partial c}\left\{\beta(c,\zeta)\left[c\,Ai(\zeta)+Ai'(\zeta)\right]\right\}
\ee
Similarly, we have
\be\label{combP2}
\beta(\zeta,c)\,Ai''(\zeta)-\,Ai'(\zeta)\beta'(c,\zeta)=\beta(c,\zeta)\left[c\,Ai'(\zeta)+Ai''(\zeta)\right]-Ai(\zeta)Ai'(\zeta)
\ee
resulting in
\be\label{P2simp}
{\cal P}^{(II)}(c,\zeta)=-\frac{\partial}{\partial c}\left\{\beta(c,\zeta)\nu(c,\zeta)\right\}
\ee
where we have denoted
\be\label{nucomb}
\nu(c,\zeta)=c\,Ai'(\zeta)+Ai''(\zeta)-\left[c\,Ai(\zeta)+Ai'(\zeta)\right]\frac{1}{2}\frac{\rho'(\zeta)}{\rho(\zeta)}
\ee

The above formulae provide exact curvature distribution in the soft-edge limit. As they are
quite complicated, it makes sense to work out several limiting cases of general interest explicitly.

\subsection{\small Ivestigating large curvature asymptotics: $c \to +\infty$ }

In this limit $\delta(c,\gamma)\equiv 0$. We can expand $\frac{1}{c^2+\tau^2}=\frac{1}{c^2}\left(1-\frac{\tau^2}{c^2}+\frac{\tau^4}{c^4}+\ldots \right)$. Substituting to (\ref{gamma}) we get
\be\label{gammalargecurv}
\gamma(c,\zeta)=\frac{1}{c^2}\sum_{k=0}^{\infty}\frac{1}{c^{2k}}Ai^{(2k)}(\zeta) =\frac{1}{c^2}\,Ai(\zeta)+\frac{1}{c^4}Ai''(\zeta)+\frac{1}{c^6}Ai''''(\zeta)+\ldots
\ee
 which implies
 \be\label{gammalargecurv1}
\gamma'(c,\zeta) =\frac{1}{c^2}\,Ai'(\zeta)+\frac{1}{c^4}Ai'''(\zeta)+\ldots
\ee
so that
\be\label{betalargecurv1}
\beta_{\gamma}(c,\zeta)\equiv c\,\gamma(c,\zeta)-\gamma'(c,\zeta)=\frac{1}{c}\sum_{k=0}^{\infty}\frac{(-1)^k}{c^{k}}Ai^{(k)}(\zeta)
\ee
Now a simple calculation gives
\be\label{P1largecurv}
{\cal P}^{(I)}_{\gamma}(c,\zeta)\equiv -c\,\beta_{\gamma}(c,\zeta) Ai(\zeta)-\beta_{\gamma}(c,\zeta) Ai'(\zeta)+Ai^2(\zeta)
\ee
\[
=\sum_{k=1}^{\infty}\frac{(-1)^k}{c^{k}}\left[Ai'(\zeta)\,Ai^{(k-1)}(\zeta)-Ai(\zeta)\,Ai^{(k)}(\zeta)\right]
\]
or explicitly, up to the terms of the order $O(c^{-4})$, using  $Ai''(\zeta)=\zeta\,Ai(\zeta)$
\be\label{P1largecurv1}
{\cal P}^{(I)}_{\gamma}(c,\zeta)\approx \frac{1}{c^2}\left(\,Ai'(\zeta)^2-\zeta\, Ai^2(\zeta) \right)+
\frac{1}{c^3}\, Ai^2(\zeta)
\ee
\[
+\frac{1}{c^4}\left\{\zeta\left[ Ai'(\zeta)^2- Ai(\zeta)A''(\zeta)\right]-Ai'(\zeta)\,Ai(\zeta)\right\}
\]
Now, rewriting (\ref{nucomb}) as
\be\label{nucomb1}
\nu(c,\zeta)=c\left(\,Ai'(\zeta)-Ai'(\zeta)\frac{1}{2}\frac{\rho'(\zeta)}{\rho(\zeta)}\right)+
\left(Ai''(\zeta)-Ai'(\zeta)\frac{1}{2}\frac{\rho'(\zeta)}{\rho(\zeta)}\right)\
\ee
 we get after straightforward manipulations
\be\label{nulargecurv}
\beta_{\gamma}(c,\zeta)\nu(c,\zeta)=\left(\,Ai'(\zeta)-Ai'(\zeta)\frac{1}{2}\frac{\rho'(\zeta)}{\rho(\zeta)}\right)Ai(\zeta)
\ee
\[
+\sum_{k=1}^{\infty}\frac{(-1)^k}{c^{k}}\left\{\,Ai^{(k)}(\zeta)\left(\,Ai'(\zeta)-Ai'(\zeta)\frac{1}{2}\frac{\rho'(\zeta)}{\rho(\zeta)}\right)
-Ai^{(k-1)}(\zeta)\left(Ai''(\zeta)-Ai'(\zeta)\frac{1}{2}\frac{\rho'(\zeta)}{\rho(\zeta)}\right)\right\}
\]
and therefore
\[
{\cal P}^{(II)}_{\gamma}(c,\zeta)=-\frac{\partial}{\partial c}\left\{\beta_{\gamma}(c,\zeta)(c,\zeta)\right\}
\]
\[
=-\sum_{k=2}^{\infty}\frac{(-1)^k(k-1)}{c^{k}}\left\{\,Ai^{(k-1)}(\zeta)\left(\,Ai'(\zeta)-Ai'(\zeta)\frac{1}{2}\frac{\rho'(\zeta)}{\rho(\zeta)}\right)\right.
\]
\[
\left.-Ai^{(k-2)}(\zeta)\left(Ai''(\zeta)-Ai'(\zeta)\frac{1}{2}\frac{\rho'(\zeta)}{\rho(\zeta)}\right)\right\}
\]
\be\label{P2largecurv1}
\approx -\frac{1}{c^2}\left(\,Ai'(\zeta)^2-\zeta\, Ai^2(\zeta) \right)-
\frac{1}{c^3}\, Ai^2(\zeta)
\ee
\[
+\frac{1}{c^4}\left(\frac{3}{2}A^2 \frac{\rho'(\zeta)}{\rho(\zeta)}-2\zeta\rho(\zeta)-3A(\zeta) A(\zeta)\right)+\ldots
\]
Adding up the two contributions we obtain the general expression
\be\label{Plargecurvgen}
{\cal P}_{\gamma}(c,\zeta)=\sum_{k=2}^{\infty}\frac{(-1)^k}{c^{k}}{\cal P}_k\,,
\ee
where
\be
{\cal P}_k=(2-k)Ai'(\zeta)Ai^{(k-1)}(\zeta)-Ai(\zeta)Ai^{(k)}(\zeta)+(k-1)Ai''(\zeta)Ai^{(k-2)}(\zeta)
\ee
\[
-(k-1)(\frac{1}{2}\frac{\rho'(\zeta)}{\rho(\zeta)}\left[Ai'(\zeta)Ai^{(k-2)}(\zeta)-Ai(\zeta)Ai^{(k-1)}(\zeta)\right]
\]
One finds that ${\cal P}_1={\cal P}_2={\cal P}_3=0$, so that the first non-vanishing term in the sum is ${\cal P}_4$, and therefore:
\be\label{Plargecurv}
{\cal P}_{\gamma}(c,\zeta)\approx \frac{1}{c^4}\left\{-2\zeta\rho(\zeta)+\frac{3}{2}A^2\frac{\rho'(\zeta)}{\rho(\zeta)}-4A(\zeta) A'(\zeta)\right\}+O(c^{-5}), \quad c\to +\infty
\ee
The above result confirms our intuition that the large curvature values occur when the two levels approach closely, hence the large-curvature
tail exponent being dictated by the level repulsion mechanism is therefore universal, see \cite{Gaspard}. For $\beta=2$ this mechanism indeed predicts ${\cal P}(c\to \infty)\sim c^{-4}$, in full agreement with (\ref{Plargecurv}).

\subsubsection{\small Towards the bulk:  $\zeta \to -\infty$ limit.}
In this limit we approach the bulk of the spectrum and it is natural to expect that the result
will match the bulk curvature distribution (\ref{DZ}).

As $\zeta=-|\zeta|$, with $|\zeta|\to \infty$, we can rewrite (\ref{gamma}) as
\[
\gamma(c,\zeta)=\frac{1}{2\pi}\int_{-\infty}^{\infty}\,d\tau\, e^{i\left(-\tau|\zeta|+\frac{\tau^3}{3}\right)}\frac{d\tau}{c^2+\tau^2}
\]
\be\label{gammaminus}
=\frac{\sqrt{|\zeta|}}{2\pi}\int_{-\infty}^{\infty} e^{i|\zeta|^{3/2}\left(-\tau+\frac{\tau^3}{3}\right)}\frac{d\tau}{c^2+|\zeta|\tau^2}
\ee
where we have changed the integration variable $\tau\to |\zeta|^{1/2}\tau$. The above integral is clearly amenable to evaluation by the saddle-point method. The s
addle-point condition is $\tau^2=1$, hence $\tau=\pm 1$. Explicit calculation gives the leading-order contribution:
\be\label{gamma0minus}
\gamma(c,-\zeta\gg 1)\approx  \frac{1}{c^2+|\zeta|}\, Ai(\zeta)\equiv \gamma_0(c,\zeta)
\ee
where we have used the asymptotic of the Airy function in the same limit:
\be\label{Airyminus}
Ai(\zeta)\approx  \frac{1}{\sqrt{\pi}|\zeta|^{1/4}} \cos{\left(\frac{2}{3}|\zeta|^{3/2}-\frac{\pi}{4}\right)}
\ee
Anticipating that we may need next to-the-leading order terms, let us consider the difference $\gamma(c,\zeta)-\gamma_0(c,\zeta)$.
Using the exact integral representation formula (\ref{Ai}) for $Ai(\zeta)$ one can show that:
\be\label{gammadif}
\gamma(c,\zeta)-\gamma_0(c,\zeta)=\frac{1}{2\pi}\frac{|\zeta|^{3/2}}{(c^2+|\zeta|)}\int_{-\infty}^{\infty}\, e^{i|\zeta|^{3/2}\left(-\tau+\frac{\tau^3}{3}\right)}\frac{(\tau^2-1)}{c^2+|\zeta|\tau^2}\,d\tau
\ee
\[
=\frac{1}{2\pi}\frac{-i}{(c^2+|\zeta|)}\int_{-\infty}^{\infty}\, \frac{1}{c^2+|\zeta|\tau^2}\,d\left\{e^{i|\zeta|^{3/2}\left(-\tau+\frac{\tau^3}{3}\right)}\right\}
\]
and performing integration by parts we arrive at the exact relation
\be\label{gammadif1}
\gamma(c,\zeta)-\gamma_0(c,\zeta)=\frac{-2i}{2\pi}\frac{|\zeta|}{(c^2+|\zeta|)}\int_{-\infty}^{\infty}\, e^{i|\zeta|^{3/2}\left(-\tau+\frac{\tau^3}{3}\right)}\frac{\tau}{(c^2+|\zeta|\tau^2)^2}\,d\tau
\ee
We again can evaluate the limit $|\zeta|\to -\infty$ in the above expression by the steepest descent method, and find
\be\label{gamma1minus}
\gamma(c,-\zeta\gg 1)-\gamma_0(c,-\zeta\gg 1)\approx \frac{2}{(c^2+|\zeta|)^3}Ai'(\zeta)\equiv \gamma_1(c,\zeta)
\ee
We conclude, that for $-\zeta\gg 1$ we can use the following approximation
\be\label{gammaminusasymp}
\gamma(c,-\zeta\gg 1)\approx   \frac{1}{c^2+|\zeta|}\, Ai(\zeta)+ \frac{2}{(c^2+|\zeta|)^3}\, Ai'(\zeta)+\ldots
\ee
(We can actually check that assuming in (\ref{gammaminusasymp}) further $c\gg |\zeta|$ and expanding reproduces the series
 (\ref{gammalargecurv}).)
Differentiating, we find to the same order
\be\label{gammaprimminusasymp}
\gamma'(c,-\zeta \gg 1)\approx   \frac{1}{c^2+|\zeta|}\, Ai'(\zeta)+ \frac{1}{(c^2+|\zeta|)^2}\, Ai(\zeta)+ \frac{2}{(c^2+|\zeta|)^3}\, Ai''(\zeta)+\ldots
\ee
Taking into account that $\delta(c,\zeta)$ is exponentially small in such a regime, we neglect it and find correspondingly
\be \label{betaminusasymp}
\beta(c,-\zeta \gg 1)\approx   \frac{c\, Ai(\zeta)-Ai'(\zeta)}{c^2+|\zeta|}-\frac{1}{(c^2+|\zeta|)^2}\, Ai(\zeta)
\ee
\[
+\frac{2}{(c^2+|\zeta|)^3}\left(Ai'(\zeta)-Ai''(\zeta)+c\,Ai'(\zeta)\right)+\ldots
\]
Substituting this to (\ref{P1simp}) gives using $\rho(\zeta)= Ai'(\zeta)^2-Ai''(\zeta)Ai(\zeta)$:
\be \label{P1minusasymp1}
{\cal P}^{(I)}(c,-\zeta\gg 1)\approx  \frac{\rho(\zeta)}{c^2+|\zeta|}+\frac{c\,Ai^2(\zeta)-Ai(\zeta)Ai'(\zeta)}{(c^2+|\zeta|)^2}+O\left(\frac{c}{(c^2+|\zeta|)^3}\right)
\ee
Similar calculation yields
\be \label{betaprimminusasymp}
\frac{\partial}{\partial c}\beta(c,-\zeta \gg 1)\approx   -\frac{ Ai(\zeta)}{c^2+|\zeta|}+\frac{2}{(c^2+|\zeta|)^2}\,\left(c\,Ai'(\zeta)- Ai''(\zeta)\right)
\ee
\[
+\frac{4c\,}{(c^2+|\zeta|)^3}\,Ai(\zeta)+\ldots
\]
and further substituting to (\ref{P2simp}) gives after straightforward but lengthy algebra
\be\label{P2minusasymp}
{\cal P}^{(II)}(c,-\zeta\gg 1)\approx -\frac{\rho(\zeta)}{c^2+|\zeta|}-\frac{c\,Ai^2(\zeta)}{(c^2+|\zeta|)^2}
\ee
\[
+\frac{1}{(c^2+|\zeta|)^2}\left\{-2\zeta\,(Ai'^2(\zeta)-\zeta Ai^2(\zeta))-3Ai(\zeta)Ai'(\zeta)+\frac{3}{2}\frac{\rho'(\zeta)}{\rho(\zeta)} Ai^2(\zeta)\right\}
\]
Adding the two contributions gives the first nonvanishing term to be
\be \label{Pminusasymp}
 {\cal P}(c,-\zeta\gg 1)\approx \frac{1}{(c^2+|\zeta|)^2}\left\{-2\zeta\rho(\zeta)+\frac{3}{2}A^2(\zeta)\frac{\rho'(\zeta)}{\rho(\zeta)}-4A(\zeta) A'(\zeta)\right\}+\ldots
 \ee
 Note that assuming $c\gg |\zeta|$ gives back exactly  (\ref{Plargecurv}). In the present case $-\zeta\gg 1$ we, however, can further use
 that $ Ai(\zeta)\sim \frac{1}{\sqrt{\pi}}\frac{1}{|\zeta|^{1/4}}, \, Ai'(\zeta)\sim -\sqrt{|\zeta|} Ai(\zeta),\,\rho(\zeta) \approx \frac{1}{\pi}\sqrt{|\zeta|},\,\, \rho'(\zeta)\approx -\frac{1}{2\pi}|\zeta|^{-1/2}\sim \frac{1}{\pi^2\rho(\zeta)}$. We then see that the leading term is the first one, and arrive at the final expression:
 \be \label{Pminusasymplead}
{\cal P}(c,-\zeta\gg 1)\approx \frac{2}{\pi}\frac{\pi^3\rho^3(\zeta)}{\left(c^2+\pi^2\rho^2(\zeta)\right)^2}\approx
\frac{2}{\pi}\frac{\kappa^3}{\left[c^2+\kappa^2\right]^2}, \quad \kappa=\pi\rho(\zeta), \,
\ee
The formula (\ref{Pminusasymplead})
precisely matches the bulk curvature distribution (\ref{DZ}), as was anticipated.

\subsubsection{\small Away from the bulk: $\zeta \to +\infty$ limit.}
Again the idea is to apply the saddle-point method for $\zeta\to \infty$. We start with the asymptotic analysis for the Airy function.
We shall see that for our goal we need actually also next to the leading order corrections to the Airy function. Let us find them from the saddle-point method. We start with the representation valid for $\zeta>0$:
 \be\label{Airyplusexec}
 Ai(\zeta)=\frac{\sqrt{\zeta}}{2\pi}\int_{-\infty}^{\infty}e^{i\zeta^{3/2}\left(\tau+\frac{\tau^3}{3}\right)}\,d\tau,\quad \zeta>0
 \ee
 The saddle-points are $\tau=\pm i$, and the relevant one is $\tau=i$ as known from the asymptotic analysis of the Airy functions(see below).
Denoting the steepest descent contour passing through $\tau=i$ as $\Gamma$, we shift to it from the original contour running along the real axis.
It is known that $\Gamma$ runs asymptotically tangent to $arg(\tau)=\pi/6$ for $\Re (\tau)\to \infty$, and $arg(\tau)=5\pi/6$ for $\Re (\tau)\to -\infty$. However, for finite $\Re{\tau}$ (in particularly, close to $\tau=i$) we can consider the contour as running actually parallel to the real axis and parametrise $\Gamma$ as $\tau=i+\frac{v}{(2\zeta)^{3/4}}$. This gives
\be\label{Airyplusexec1}
 Ai(\zeta)=\frac{1}{2\sqrt{\pi} \zeta^{1/4}}e^{-\frac{2}{3}\zeta^{3/2}} J\left(a\right), \quad a=(2\zeta)^{3/4}, \quad
 J(a)= \int_{-\infty}^{\infty}e^{-\frac{v^2}{2}+i\frac{v^3}{a}}\,\frac{dv}{\sqrt{2\pi}},
 \ee
 The corresponding leading-order asymptotic for the Airy functions is obtained by replacing $J\left(a\right)=1$:
 \be\label{Airypluslead}
 Ai(\zeta)\approx  Ai_0(\zeta)=\frac{1}{2\sqrt{\pi}\zeta^{1/4}}e^{-\frac{2}{3}\zeta^{3/2}},
 \ee
 and a similar calculation for $Bi$ gives
  \be\label{Birypluslead}
  Bi(\zeta)\approx Bi_0(\zeta)=\frac{1}{\sqrt{\pi}\zeta^{1/4}}e^{\frac{2}{3}\zeta^{3/2}},\quad \zeta\gg 1\,.
  \ee
 These relations imply to the leading order:
 \be\label{Airyplusdirlead}
  Ai'(\zeta)\approx -\sqrt{\zeta}\,  Ai_0(\zeta), \,\, Bi'(\zeta)\approx \sqrt{\zeta}\,  Bi_0(\zeta),\quad \zeta\gg 1
 \ee
 which indeed gives the correct Wronskian, cf. (\ref{check}).
However, for our present goals we will need to account for subleading terms as well. To see this fact we can consider
the mean density $\rho(\zeta)= Ai'(\zeta)^2-\zeta\,Ai^2(\zeta)$ featuring in the calculation. Using (\ref{Airypluslead},\ref{Airyplusdirlead}) immediately shows that the leading order vanishes, and we indeed need to go to the subleading order by expanding $J\left(a\gg 1\right)=1-\frac{5!!}{2!}\frac{1}{a^2}+O(a^{-4})$. This gives for the Airy function:
\be\label{Airyplussublead}
 Ai(\zeta)\approx  Ai_0(\zeta)\left(1-\frac{15}{16}\frac{1}{\zeta^{3/2}}+O(\zeta^{-3})\right), \quad \zeta\gg 1
 \ee
 After differentiation, using (\ref{Airypluslead})
\be\label{Airydirplussublead}
 Ai'(\zeta)\approx  -Ai_0(\zeta)\,\zeta^{1/2}\left(1-\frac{11}{16}\frac{1}{\zeta^{3/2}}+O(\zeta^{-3})\right), \quad
  \quad \zeta\gg 1
 \ee
With this expressions we find using $\rho'(\zeta)=-Ai^2(\zeta)$ that
\be\label{denpluslead}
\rho(\zeta)\approx \frac{1}{2\sqrt{\zeta}}\,\left[Ai_0(\zeta)\right]^2\left(1+O(\zeta^{-3/2})\right), \quad
\frac{\rho'(\zeta)}{2\rho(\zeta)}\approx -\sqrt{\zeta}\left(1+\frac{b}{16\zeta^{3/2}}\right)
\ee
 with some yet unspecified coefficient $b$.
 Consider now the combination (\ref{nucomb}), conveniently re-arranged
\be\label{nucomb11}
\nu(c,\zeta)=c\left[Ai'(\zeta)-Ai(\zeta)\frac{1}{2}\frac{\rho'(\zeta)}{\rho(\zeta)}\right]+\zeta Ai(\zeta)-Ai'(\zeta)\frac{1}{2}\frac{\rho'(\zeta)}{\rho(\zeta)}
\ee
We have
\[
Ai'(\zeta)-Ai(\zeta)\frac{1}{2}\frac{\rho'(\zeta)}{\rho(\zeta)}\approx Ai_0(\zeta)\frac{1}{16\zeta}(b-4) ,
\]

\[
\zeta Ai(\zeta)-Ai'(\zeta)\frac{1}{2}\frac{\rho'(\zeta)}{\rho(\zeta)}\approx Ai_0(\zeta)\frac{1}{16\sqrt{\zeta}}(b+4)
\]
so that
\be\label{nucomb2}
\nu(c,\zeta)\approx Ai_0(\zeta)\frac{1}{16\zeta}\left[c(b-4)-\sqrt{\zeta}(b+4)\right]
\ee
Therefore, we have asymptotically to the leading order
\be\label{P2asymp}
{\cal P}^{(II)}(c,\zeta)\approx -Ai_0(\zeta)\frac{1}{16\zeta}\frac{\partial}{\partial c}\left\{\beta(c,\zeta)\left[c(b-4)-\sqrt{\zeta}(b+4)\right]\right\}
\ee
and to the same order
\[
{\cal P}^{(I)}(c,\zeta)=-\beta(c,\zeta)\left[c\,Ai(\zeta)+Ai'(\zeta)\right]+Ai^2(\zeta)
\]
\be\label{P1asymp}
\approx \beta(c,\zeta)Ai_0(\zeta)(\sqrt{\zeta}-c)+Ai^2_0(\zeta)
\ee

It remains to find asymptotic for $\beta(c,\zeta)$. To this end,
 we perform a similar analysis for $\gamma(c,\zeta)$. We write
\be\label{gammaplus}
\gamma(c,\zeta)=\frac{\sqrt{\zeta}}{2\pi}\int_{-\infty}^{\infty}\,d\tau\, e^{i\zeta^{3/2}\left(\tau+\frac{\tau^3}{3}\right)}\frac{d\tau}{c^2+\zeta\tau^2}
\ee
\[
=\frac{1}{4\pi i c}\left[\int_{-\infty}^{\infty}\,d\tau\, e^{i\zeta^{3/2}\left(\tau+\frac{\tau^3}{3}\right)}\frac{d\tau}{\tau-ic/\sqrt{\zeta}}-\int_{-\infty}^{\infty}\,d\tau\, e^{i\zeta^{3/2}\left(\tau+\frac{\tau^3}{3}\right)}\frac{d\tau}{\tau+ic/\sqrt{\zeta}}\right]
\]  For $|c|<\sqrt{\zeta}$ deforming contour from the real line to the steepest descent contour $\Gamma$ incurs the contribution from the pole at $\tau=i|c|/\sqrt{\zeta}$,
  whereas for $|c|>\sqrt{\zeta}$ the pole is above $\Gamma$. We therefore have the exact identity:
  \be\label{gammaplus1}
\gamma(c,\zeta>0)=\gamma_{p}(c,\zeta>0)+\gamma_{\Gamma}(c,\zeta>0)
\ee
where we denoted the pole contribution
\be\label{gamma0}
\gamma_{p}(c,\zeta>0)=\frac{1}{2|c|}e^{-\zeta|c|+\frac{1}{3}|c|^3}\theta(\sqrt{\zeta}-|c|)
\ee
and
\be\label{gammaGamma}
\gamma_{\Gamma}(c,\zeta>0)=\frac{\sqrt{\zeta}}{2\pi}\int_{\Gamma}\,d\tau\, e^{i\zeta^{3/2}\left(\tau+\frac{\tau^3}{3}\right)}\frac{d\tau}{c^2+\zeta\tau^2}
\ee
Performing the integral by the steepest descent method gives ( using (\ref{Airypluslead}))
\be\label{gammaGammaasy}
\gamma_{\Gamma}(c,\zeta>0)\approx \frac{1}{c^2-\zeta}\,Ai_0(\zeta), \quad \zeta\gg 1, |c|\ne \sqrt{\zeta}
\ee
It turns out that for $|c|<\sqrt{\zeta}$ the term $\gamma_{p}$ gives much larger contribution than $\gamma_{\Gamma}$.
To verify this we rescale $c=\tilde{c}\sqrt{\zeta}$ and consider $\tilde{c}$ to be of the order of unity. For
$\gamma_{p}$ we arrive at the expression:
\be\label{gammapscaled}
\gamma_{p}(c,\zeta>0)=\frac{1}{2|\tilde{c}|\zeta^{1/2}}e^{-\zeta^{3/2}\left(f(|c|)\right)}\theta(1-|\tilde{c}|), \quad f(u)=u-\frac{1}{3}u^3
\ee
As $f'(u)=1-u^2 > 0$ for $u\in[0,1)$ we see that $f(u)<f(1)=2/3$, hence   $\gamma_{p}(c,\zeta>0)\gg \frac{1}{\zeta^{1/2}} e^{-\frac{2}{3}\zeta^{3/2}}$. At the same time using the asymptotics of the Airy function $\gamma_{\Gamma}\sim \frac{1}{2\sqrt{\pi}\zeta^{5/4}} e^{-\frac{2}{3}\zeta^{3/2}}<\frac{1}{\zeta^{1/2}} e^{-\frac{2}{3}\zeta^{3/2}}\ll \gamma_p $, proving the statement.

Thus for $|c|<\sqrt{\zeta}$ we can approximate $\gamma(c,\zeta)\approx \gamma_{p}(c,\zeta)$, which upon differentiation gives  \[\gamma'_p(c,\zeta>0)= -\frac{1}{2}e^{-\zeta|c|+\frac{1}{3}|c|^3}\theta(\sqrt{\zeta}-|c|)\] and further
$$c\,\gamma_p(c,\zeta>0)-\gamma_p'(c,\zeta>0)\approx e^{-\zeta\,c+\frac{1}{3}c^3}\,\theta(c)\theta(\sqrt{\zeta}-c)$$
and finally adding (\ref{delta1}) we obtain
\be\label{betap}
\beta(c,\zeta>0)\approx \beta_p(c,\zeta>0)\equiv c\,\gamma_p(c,\zeta>0)-\gamma_p'(c,\zeta>0)+\delta(c,\zeta)= e^{-\zeta\,c+\frac{1}{3}c^3}\theta(\sqrt{\zeta}-c)
\ee
which is the leading approximation ( exponentially dominant) in the whole domain $c\in (-\infty, \sqrt{\zeta})$.

It is useful to check that such a precision is sufficient for the relation (\ref{check}) to hold:
\be\label{check1}
\int_{-\infty}^{\infty}\, \beta(c,\zeta)\,dc=\int_{-\infty}^{\sqrt{\zeta}}\,e^{-\zeta\,c+\frac{1}{3}c^3}\,dc\approx
\pi Bi(\zeta)
\ee
which is checked by performing the integral (after change $c\to \sqrt{\zeta}\,c$ ) by the steepest descent method, and using
asymptotic (\ref{Birypluslead}) for $Bi(\zeta)$.

Substituting (\ref{betap}) to (\ref{P1asymp}) gives the leading order expression:
\be\label{P1pasymp}
{\cal P}^{(I)}_p(c,\zeta)\approx Ai_0(\zeta)\,e^{-\zeta\,c+\frac{1}{3}c^3}(\sqrt{\zeta}-c), \quad c\in (-\infty, \sqrt{\zeta})
\ee
Differentiation in the domain $c\in (-\infty, \sqrt{\zeta})$ gives
\be\label{betapprim}
\frac{\partial}{\partial c}\beta_p(c,\zeta)= (c^2-\zeta)\,\beta_p(c,\zeta), \quad c\in (-\infty, \sqrt{\zeta})
\ee
which finally shows that to the leading order
\be\label{P2pasymp}
{\cal P}^{(II)}_p(c,\zeta)\approx Ai_0(\zeta)\,e^{-\zeta\,c+\frac{1}{3}c^3}(\zeta-c^2)\frac{1}{16\zeta} \left[c(b-4)-\sqrt{\zeta}(b+4)\right], \quad c\in (-\infty, \sqrt{\zeta})
\ee
Adding together we get the full curvature p.d.f. for $\zeta\gg 1$, and $c\in (-\infty, \sqrt{\zeta})$:
\be\label{Ppasymp}
{\cal P}_p(c,\zeta)\approx Ai_0(\zeta)\,e^{-\zeta\,c+\frac{1}{3}c^3}(\sqrt{\zeta}-c)
\left\{1+\frac{1}{16}(1+\frac{c}{\sqrt{\zeta}}) \left[\frac{c}{\sqrt{\zeta}}(b-4)-(b+4)\right]\right\}
\ee
To understand the structure of the above expression notice that
\be\label{mainfactor}
 Ai_0(\zeta)\,e^{-\zeta\,c+\frac{1}{3}c^3}=\frac{1}{2\sqrt{\pi} \zeta^{1/4}}e^{\frac{1}{3}c^3-c\zeta-\frac{2}{3}\zeta^{3/2}}=\frac{1}{2\sqrt{\pi} \zeta^{1/4}}e^{-\frac{1}{3}(2\sqrt{\zeta}-c)(c+\sqrt{\zeta})^2}
\ee
so that the curvature p.d.f. has a very sharp maximum (with the height $\zeta^{1/4}$ and widths $\zeta^{-1/4}$)  around $c=-\sqrt{\zeta}$.
 Introduce correspondingly the new random variable $x=(c+\sqrt{\zeta})\sqrt{2}\zeta^{1/4}$, so that $c=\frac{x}{\sqrt{2}\zeta^{1/4}}-\sqrt{\zeta}$.
\be\label{mainfactor1}
 Ai_0(\zeta)\,e^{-\zeta\,c+\frac{1}{3}c^3}=\frac{1}{2\sqrt{\pi} \zeta^{1/4}}e^{-\frac{x^2}{2}+\frac{1}{6\sqrt{2}}\frac{x^3}{\zeta^{3/4}}}
\ee
Remembering that
the p.d.f. of $x$ acquires the extra Jacobian factor $\sqrt{2}\zeta^{1/4}$, we see that the probability density
\be\label{P1x}
{\cal P}_p(x,\zeta\gg 1)\approx \frac{1}{\sqrt{2\pi}}e^{-\frac{x^2}{2}}e^{\frac{1}{6\sqrt{2}}\frac{x^3}{\zeta^{3/4}}}\left(1-\frac{x}{2\sqrt{2}\zeta^{3/4}}\right)
\left\{1+O\left(\frac{x}{\zeta^{3/4}}\right)\right\}\
\ee
tends to the standard gaussian distribution $\frac{1}{\sqrt{2\pi}}e^{-\frac{x^2}{2}}$ for fixed $x$ and $\zeta\to \infty$.

Now it remains to consider the case $c>\sqrt{\zeta}$. Then the only contribution comes from the term $\gamma_{\Gamma}(\zeta)$.
Using precisely the same method as in $\zeta \to -\infty$ limit we can arrive at the full analogue of the exact identity (\ref{gammadif1}):
\be\label{gammadif1A}
\gamma_{\Gamma}(c,\zeta)-\frac{1}{c^2-\zeta}\,Ai(\zeta)=\frac{2i}{2\pi}\frac{\zeta}{(c^2-\zeta)}\int_{-\infty}^{\infty}\, e^{i\zeta^{3/2}\left(\tau+\frac{\tau^3}{3}\right)}\frac{\tau}{(c^2-\zeta\tau^2)^2}\,d\tau
\ee
and performing the integral by steepest descent further see
\be\label{Gammaapprox}
\gamma_{\Gamma}(c,\zeta)\approx \frac{1}{c^2-\zeta}\,Ai(\zeta)+\frac{2}{(c^2-\zeta)^2}\,Ai(\zeta)+\ldots
\ee
Noticing that $|\zeta|=-\zeta$ for $\zeta<0$, we see that (\ref{Gammaapprox}) is precisely the same as (\ref{gammaminusasymp}), hence
the curvature p.d.f. will be again given by analogue of (\ref{Pminusasymp}):
\be \label{Plusasymp}
 {\cal P}(c>\sqrt{\zeta}\gg 1)\approx \frac{1}{(c^2-\zeta)^2}\left\{-2\zeta\rho(\zeta)+\frac{3}{2}A^2\frac{\rho'(\zeta)}{\rho(\zeta)}-4A(\zeta) A'(\zeta)\right\}+\ldots
 \ee
Using the asymptotic formulae (\ref{Airyplussublead}), (\ref{denpluslead}) we find that the leading term in (\ref{Plusasymp})
cancels, and the result is of the order of $Ai_0^2(\zeta)\zeta^{-3/2}\sim \frac{\rho}{\zeta} $, or explicitly
\be \label{Plusasympexp}
 {\cal P}(c>\sqrt{\zeta}\gg 1)\approx \frac{2B}{(c^2-\zeta)^2}\frac{\rho(\zeta)}{\zeta} \approx \frac{B}{(c^2-\zeta)^2} \frac{1}{4\pi\zeta^2}\,e^{-\frac{4}{3}\zeta^{3/2}},
 \ee
where the constant $B$ of the order of unity is left undetermined.

The crossover between the two regimes $c<\sqrt{\zeta}$ and $c>\sqrt{\zeta}$ happens over the domain $|c-\sqrt{\zeta}|\sim \frac{1}{\zeta^{1/4}}$. Indeed, take $c=\sqrt{\zeta}+ \frac{\beta}{2\zeta^{1/4}}$, where $\beta$ is of the order of unity. Substituting this for $\beta>0$ into $\gamma_{\Gamma}$ from (\ref{gammaGammaasy}) and using asymptotic for Airy function at $\zeta\gg 1$ we see that
\be
\gamma_{\Gamma}\approx  \frac{1}{2\sqrt{\zeta}(c-\zeta)}\frac{}{2\sqrt{\pi}\zeta^{1/4}}\,e^{-\frac{2}{3}\zeta^{3/2}}\approx \frac{1}{2\sqrt{\pi\zeta}\beta}\,e^{-\frac{2}{3}\zeta^{3/2}}
\ee
On the other hand, the pole contribution (\ref{gamma0}) for $c\approx \sqrt{\zeta}$ approaches the value
\be
\gamma_{p}\approx  \frac{1}{2\sqrt{\zeta}}\,e^{-\frac{2}{3}\zeta^{3/2}}
\ee
that is the two contributions are of the same order. It would be interesting to find the exact crossover expression for arbitrary fixed $\beta$,
and we leave it for further investigation.

{\bf Acknowledgements}\\
The author is grateful to the organizers of the 5th Workshop on Quantum Chaos and Localization Phenomena, May 20-22, 2011, Warsaw, Poland
 for supporting his participation.

\section*{\small Appendix B: Bulk asymptotics of Hermite polynomials, Cauchy transforms, and the kernels involved.}

  As is well-known ( see e.g. \cite{mylec}) the Hermite polynomials $p_k(x)$ have the following integral representation:
 \be\label{polybasic}
 p_{N+n}(x)=\sqrt{\frac{N}{2\pi}} \left[(-i)^{N+n}I_{N+n}(x)+i^{N+n}\overline{I_{N+n}(x)}\right]
 \ee
 where
 \be\label{I}
  \quad I_{N+n}(x)=\int_0^{\infty}\,dq\,q^{N+n}
 e^{-\frac{N}{2}(q-ix)^2}=e^{N\frac{x^2}{2}}\int_0^{\infty}\,dq\,q^{N+n}
 e^{-\frac{N}{2}q^2+iNxq }
\ee
For real $|x|<2$ we can use parametrization denoting $x=2\cos{\phi}$ and by the steepest descent method (s.p. at $q=ie^{-i\phi}$) find  \cite{mylec}
 the following large$-N$ asymptotic behaviour:
\be\label{bulkas}
I_{N+n}(x)=i^{N+n}\sqrt{\frac{\pi}{N \sin{\phi}}}\, e^{\frac{N}{2}\cos{2\phi}}\,e^{-i\left( n+\frac{1}{2}\right)\phi+i\frac{\pi}{4}-i\,N\theta(\phi)}, \quad \theta(\phi)=\phi-\frac{1}{2}\sin{(2\phi)}
\ee
so that
\be\label{bulkaspolyn}
p_{N}(x)\approx \sqrt{\frac{2}{ \sin{\phi}}}\, e^{\frac{N}{2}\cos{2\phi}}\cos{\left(\frac{1}{2}\phi-\frac{\pi}{4}+N\theta(\phi)\right)}
\ee
Differentiating over $x$ , picking up the leading terms proportional to $N$, and using $\frac{d\phi}{dx}=-\frac{1}{2\sin{\phi}}$ and $\frac{d\theta(\phi)}{d\phi}=2\sin^2{(\phi)}$ we get
\[
\frac{d}{dx}p_{N}(x)\approx  -\frac{1}{2\sin{\phi}}\sqrt{\frac{2}{\sin{\phi}}}\, e^{\frac{N}{2}\cos{2\phi}} \, N \times
\]
\[
\times \left[-\sin({2\phi)}\cos{\left(\frac{1}{2}\phi-\frac{\pi}{4}+N\theta(\phi)\right)}-2\sin^2{(\phi)}\sin{\left(\frac{1}{2}\phi-\frac{\pi}{4}+N\theta(\phi)\right)}\right]
\]
\be \label{d1}
=Np_{N-1}(x)
\ee
which in fact is easy to show to be the {\it exact} relation for Hermite polynomials. Similarly

\be\label{d2}
p_{N-1}(x)\approx \sqrt{\frac{2}{ \sin{\phi}}}\, e^{\frac{N}{2}\cos{2\phi}}\cos{\left(-\frac{1}{2}\phi-\frac{\pi}{4}+N\theta(\phi)\right)},
\quad \frac{d}{dx}p_{N-1}(x)\approx Np_{N-2}(x)
\ee

This shows that the kernel $W_1(x,x)$ from (\ref{9b}) can be written as
\be\label{ker1}
W_1(x,x)= N\left[p_{N-1}(x)p_{N-1}(x)-p_{N}(x)p_{N-2}(x)\right]\approx 2N\sin{\phi}\,e^{N \cos{2\phi}}
\ee
where we have used the identity $\cos^2(A)-\cos{(A+\phi)}\cos{(A-\phi)}=\sin^2{\phi}$ for any $A$. This implies
\be\label{ker2}
W_2(x,x)=\frac{d}{dx} W_1(x,x)\approx (-2N\sin{(2\phi)})(-\frac{1}{2\sin{\phi}})W_1(x,x)=2N\cos{\phi}\,W_1(x,x)
\ee

Finally, recalling the mean eigenvalue density $\rho(x)=\frac{1}{2\pi}\sqrt{4-x^2}$ we recover (\ref{kerker}).

Now, denote the Cauchy transforms of the above $p_k(x)$ as $h_k(x)\equiv f_k(x)$. As is shown in \cite{mylec}
for real $x$ and $\epsilon$ holds the exact integral representation
\be\label{Cauchytr}
f_{N+n}(x+i\epsilon)=-(-i)^{N+n}s_{\epsilon}^{N+n-1}\sqrt{\frac{N}{2\pi}}\int_0^{\infty}e^{-\frac{N}{2}q^2
+is_{\epsilon}(x+i\epsilon) Nq}q^{N+n}\,dq
\ee
where we denoted $s_{\epsilon}=\mbox{\small sign}(\epsilon)$.
Comparing with (\ref{I}) we conclude:
\be\label{Cauchytr1}
f_{N+n}(x+i\epsilon)=-(-i)^{N+n}s_{\epsilon}^{N+n-1}\sqrt{\frac{N}{2\pi}}
e^{-\frac{N}{2}(xs_{\epsilon}+i|\epsilon|)^2} I_{N+n}\left(x s_{\epsilon}+i|\epsilon|\right)
\ee
In our applications we will need $\epsilon=i\frac{\omega }{N}y$ (see eqs. (\ref{bulkcurv}) and (\ref{6b},\ref{F1},\ref{F2})) when considering the large-$N$ asymptotics. To this end, the formula (\ref{bulkas}) obviously
implies in the large$-N$ limit to the leading order, with real $\zeta=O(1)$
\be
I_{N+n}\left(x+\frac{\zeta}{N}\right)\approx I_{N+n}(x)\, e^{-N\sin{2\phi}\,\delta\phi-iN\theta'(\phi)\,\delta\phi}, \quad
\ee
where
\[
\delta{\phi}=\frac{d\phi}{dx}\,\delta{x}=-\frac{1}{2\sin{\phi}}\frac{\zeta}{N}, \quad \theta'(\phi)=1-\cos{(2\phi)}=2\sin^2{\phi}\,.
\]
Taking into account $\sin{\phi}=\frac{1}{2}\sqrt{4-x^2}\equiv \pi\rho(x)$ we rewrite the above as
\be\label{bulkas2}
I_{N+n}(x+\frac{\zeta}{N})\approx I_{N+n}(x)\, e^{\zeta(\cos{\phi}+i\sin{\phi})}\equiv I_{N+n}(x)\,e^{\zeta (\frac{x}{2}+i\pi\rho(x))}
\ee
which retains its validity when we replace real $\zeta$ with purely imaginary $\zeta=i|\omega|\, y$
as is needed in our application (remember $y>0$ by definition).
\be\label{bulkas2a}
I_{N+n}\left(x\,s_{\omega}+i\frac{|\omega|}{N}y\right)\approx \,e^{i|\omega| y (\frac{x s_{\omega}}{2}+i\pi\rho(x))}I_{N+n}(xs_{\omega})
\ee

Combining this with (\ref{Cauchytr1})
gives needed leading-order asymptotic of the Cauchy-transform:
\be\label{Cauchytr1a}
f_{N+n}\left(x+i\frac{\omega }{N}y\right)\approx -(-i)^{N+n}s_{\omega}^{N+n-1}\sqrt{\frac{N}{2\pi}}
e^{-\frac{N}{2}x^2} e^{-i\omega\frac{x}{2}y-\pi \rho(x)|\omega|y} I_{N+n}\left(x s_{\omega}\right)
\ee
Now we can calculate the asymptotic of the kernel $F_1\left(x,x+i\frac{\omega}{N}y\right)$ from (\ref{F1}).
Actually, it is more convenient to consider a slightly more general case:
\be\label{F1a}
F_1\left(x+\frac{\eta}{N},x+i\frac{\omega}{N}y\right)=f_{N}\left(x+i\frac{\omega}{N}y\right)p_{N-1}\left(x+\frac{\eta}{N}\right)-
f_{N-1}\left(x+i\frac{\omega}{N}y\right)p_{N}\left(x+\frac{\eta}{N}\right)
\ee
\[
=-(-i)^{N-1}s_{\omega}^{N-2}e^{-\frac{N}{2}x^2} e^{-i\omega\frac{x}{2}y-\pi \rho(x)|\omega|y}\left[-is_{\omega}I_{N}(xs_{\omega})p_{N-1}\left(x+\frac{\eta}{N}\right)-
I_{N-1}(xs_{\omega})p_{N}\left(x+\frac{\eta}{N}\right)\right]
\]

Assume first $\omega>0$, then
\[
\left[-is_{\omega}I_{N}(xs_{\omega})p_{N-1}\left(x+\frac{\eta}{N}\right)-
I_{N-1}(xs_{\omega})p_{N}\left(x+\frac{\eta}{N}\right)\right]=
\]
\[
\left[-iI_{N}(x)p_{N-1}\left(x+\frac{\eta}{N}\right)-
I_{N-1}(x)p_{N}\left(x+\frac{\eta}{N}\right)\right]
\]
or, using (\ref{polybasic}), (\ref{bulkas2}) and the relation $I_{N}(-x)=\overline{I_{N}(x)}$ for any real $x$, we have (up to the overall factor $\sqrt{\frac{N}{2\pi}}$)
\[
\propto e^{\frac{x}{2}\eta}\left[-iI_{N}(x)\left((-i)^{N-1}I_{N-1}(x)e^{i\pi\rho(x)\eta}+c.c.\right)-
I_{N-1}(x)\left((-i)^{N}I_{N}(x)e^{i\pi\rho(x)\eta}+c.c\right)\right]
\]
\[
=-i^N e^{\frac{x}{2}\eta+i\pi\rho(x)\eta}\left[I_{N}(x)I_{N-1}(-x)+I_{N}(-x)I_{N-1}(x)\right]
\]
Now assume $\omega<0$, then similar calculation gives:
\[
\left[-is_{\omega}I_{N}(xs_{\omega})p_{N-1}\left(x+\frac{\eta}{N}\right)-
I_{N-1}(xs_{\omega})p_{N}\left(x+\frac{\eta}{N}\right)\right]
\]
\[
=\left[iI_{N}(-x)p_{N-1}\left(x+\frac{\eta}{N}\right)-
I_{N-1}(-x)p_{N}\left(x+\frac{\eta}{N}\right)\right]
\]
\[
=-(-i)^Ne^{\frac{x}{2}\eta-i\pi\rho(x)\eta}\sqrt{\frac{N}{2\pi}} \left[I_{N}(x)I_{N-1}(-x)+I_{N}(-x)I_{N-1}(x)\right]
\]

Thus we can conclude that
\be
\left[-is_{\omega}I_{N}(xs_{\omega})p_{N-1}\left(x+\frac{\eta}{N}\right)-
I_{N-1}(xs_{\omega})p_{N}\left(x+\frac{\eta}{N}\right)\right]=
\ee
\[
-(s_{\omega}i)^N\sqrt{\frac{N}{2\pi}} e^{\frac{x}{2}\eta-i\pi\rho(x)\eta s_{\omega}}\left[I_{N}(x)\overline{I_{N-1}(x)}+\overline{I_{N}(x)}I_{N-1}(x)\right]
\]
which gives for the leading-order kernel asymptotics:
\[
F_1\left(x+\frac{\eta}{N},x+i\frac{\omega}{N}y\right)
\]
\be\label{F1b}
\approx i\frac{N}{2\pi}\,e^{-\frac{N}{2}x^2} e^{\frac{x}{2}(\eta-i\omega y)-\pi \rho(x)|\omega|y-i\pi \rho(x)\eta} \left[I_{N}(x)\overline{I_{N-1}(x)}+\overline{I_{N}(x)}I_{N-1}(x)\right]
\ee
Using (\ref{bulkas}) we find $I_{N}(x)\overline{I_{N-1}(x)}=i\frac{\pi}{N\sin{\phi}}e^{N\cos{(2\phi)}-i\phi}$, so that finally we find for the kernel:
\be\label{F1c}
F_1\left(x,x+i\frac{\omega}{N}y\right)\approx i\,e^{-\frac{N}{2}x^2} e^{-i\omega\frac{x}{2}y-\pi \rho(x)|\omega|y} e^{N\cos{2\phi}}=i \,e^{-N} e^{-i\omega\frac{x}{2}y-\pi \rho(x)|\omega|y}
\ee
equivalent to the  equation (\ref{F1as}).

Finally,  the kernel $F_2\left(x,x+i\frac{\omega}{N}y\right)$ is obtained by differentiating (\ref{F1b}) over $\eta$, setting $\eta=0$,  and multiplying by factor $N$.
which yields the relation
\be\label{F2c}
F_2\left(x,x+i\frac{\omega}{N}y\right)\approx N\,\left[\frac{x}{2}-is_{\omega}\pi\rho(x)\right]\,F_1\left(x,x+i\frac{\omega}{N}y\right)
\ee
which is equivalent to (\ref{F2as}).

\section*{\small Appendix B: "Soft edge" asymptotics of Hermite polynomials, Cauchy transforms, and the kernels involved.}
We start again with
\be\label{J}
  \quad I_{N+n}(x)=e^{N\frac{x^2}{2}}J_{N+n}(x), \quad J_{N+n}(x)=\int_0^{\infty}\,dq\,q^{N+n}
 e^{-\frac{N}{2}q^2+iNxq }
\ee
and replace the contour with the sum of two contours $[0,i]\bigcup [i,i+\infty]$, so that, correspondingly, $J_{N+n}(x)=J_{N+n}^{(I)}(x)+J_{N+n}^{(II)}(x)$. In the first contour we parametrize $q=ip,\, p\in[0,1]$, so that
\be\label{J1}
   J_{N+n}^{(I)}(x)=i^{N+n+1}\int_0^{1}\,dp\,p^{N+n}
 e^{\frac{N}{2}p^2-Nxp}
\ee
and in the second contour we put $q=i+t, \forall t>0$, so that
\be\label{J2}
   J_{N+n}^{(II)}(x)=\int_0^{\infty}\,dt\,(i+t)^{N+n}
 e^{-\frac{N}{2}(t+i)^2+iNx(t+i)}
\ee
We will be interested in the regime $x=2+\frac{\zeta}{N^{2/3}}$, with $\zeta$ of the order of unity, and $N\gg 1$. Let us start with rewriting (\ref{J2}) as
\be\label{J2a}
   J_{N+n}^{(II)}(x)=e^{-\frac{3}{2}N}\int_0^{\infty}\,dt\,(i+t)^{N+n}e^{iN^{\frac{1}{3}}(i+t)\zeta}
 e^{-N{\cal L}(t)}, \quad {\cal L}(t)=\frac{t^2}{2}-it-\ln{(i+t)}
\ee
The saddle-point equation is $\frac{d{\cal L}}{dt}=0=t-i-\frac{1}{t+i}$, which has the only solution $t=0$. Expanding for $t\ll 1$ gives
${\cal L}(t)\approx -\ln{i}-i\frac{t^3}{3}+O(t^4)$. Introducing the scaled variable: $t=\frac{\tau}{N^{1/3}}$ we easily find
to the leading order
\be\label{J2ass}
   J_{N+n}^{(II)}(x)\approx \frac{i^{N+n}}{N^{1/3}}\,e^{-\frac{3}{2}N}\,e^{-N^{\frac{1}{3}}\zeta}a_1(\zeta), \quad a_1(\zeta)= \int_0^{\infty}\,d\tau\,e^{i\tau\zeta+i\frac{\tau^3}{3}}
\ee
 Similarly, in the integral (\ref{J1}) we make the substitution $p=1-\frac{\tau}{N^{1/3}}$, with $\tau\in[0,N^{1/3}\to \infty]$ and after replacing $x=2+\frac{\zeta}{N^{2/3}}$
 and expanding for $\tau\ll N^{1/3}$ find to the leading order
 \be\label{J1ass}
   J_{N+n}^{(I)}(x)\approx \frac{i^{N+n+1}}{N^{1/3}}\,e^{-\frac{3}{2}N}\,e^{-N^{\frac{1}{3}}\zeta}a_2(\zeta), \quad a_2(\zeta)=\int_0^{\infty}\,d\tau\,e^{\tau\zeta-\frac{\tau^3}{3}}
\ee
 Combining, we finally have for  $x=2+\frac{\zeta}{N^{2/3}}$
 \be\label{Iassysoft}
   I_{N+n}(x)\approx \frac{i^{N+n}}{N^{1/3}}\,e^{\frac{1}{2}N}\,e^{N^{\frac{1}{3}}\zeta}\, a(\zeta), \quad a(\zeta)=a_1(\zeta)+ia_2(\zeta), \quad \forall \zeta\in\mathbb{C}
\ee
{\bf Note:} Introduce the two functions:
\be\label{Ai}
Ai(\zeta)=\frac{1}{\pi}\, \int_0^{\infty}\,d\tau\,\cos{\left(\tau\zeta+\frac{\tau^3}{3}\right)}
\ee
and
\be\label{Bi}
Bi(\zeta)=\frac{1}{\pi}\, \int_0^{\infty}\,d\tau\,\sin{\left(\tau\zeta+\frac{\tau^3}{3}\right)}+\frac{1}{\pi}\int_0^{\infty}\,d\tau\,e^{\tau\zeta-\frac{\tau^3}{3}}
\ee
which are the two standard linear independent solutions of the {\sf Airy} equation $f''(\zeta)-\zeta\,f(\zeta)=0$. Obviously, $a(\zeta)=\pi
\left[Ai(\zeta)+iBi(\zeta)\right]$.

 After substitution to (\ref{polybasic}) this yields (for real $\zeta$ and $\forall n\ll N$) to the leading order :
  \be\label{passyledsoft}
   p_{N+n}\left(2+\frac{\zeta}{N^{2/3}}\right)\approx \sqrt{{2\pi}} N^{1/6}\,e^{\frac{1}{2}N}\,e^{N^{\frac{1}{3}}\zeta}Ai(\zeta) ,
\ee
We see that to this order there is no dependence on $n$, which will result in vanishing of the corresponding kernel.
To find beyond-the-leading order corrections we will use the exact recursion: $p_{N-1}(x)=\frac{1}{N}\frac{d}{dx}p_N(x)$.
In the "soft edge" scaling regime we, correspondingly, have
\be\label{edgerecur1}
p_{N-1}\left(2+\frac{\zeta}{N^{2/3}}\right)=\frac{1}{N^{1/3}}\frac{d}{d\zeta}p_N\left(2+\frac{\zeta}{N^{2/3}}\right),
\ee
\be\label{edgerecur2}
p_{N-2}\left(2+\frac{\zeta}{N^{2/3}}\right)=\frac{1}{N^{2/3}}\frac{d^2}{d\zeta^2}p_N\left(2+\frac{\zeta}{N^{2/3}}\right)
\ee
which results in
\[
   p_{N}\left(2+\frac{\zeta}{N^{2/3}}\right)\approx A_N\,e^{N^{\frac{1}{3}}\zeta}Ai(\zeta),\quad A_N=\sqrt{2\pi}\,N^{1/6}\,e^{\frac{1}{2}N}
\]
\be\label{passycorrectsoft}
   p_{N-1}\left(2+\frac{\zeta}{N^{2/3}}\right)\approx A_N\,e^{N^{\frac{1}{3}}\zeta}\,\left[ Ai(\zeta)+\frac{1}{N^{1/3}}Ai'(\zeta)\right],
\ee
\[
 p_{N-2}\left(2+\frac{\zeta}{N^{2/3}}\right)\approx A_N\,e^{N^{\frac{1}{3}}\zeta}\,\left[ Ai(\zeta)+\frac{2}{N^{1/3}}Ai'(\zeta)+\frac{1}{N^{2/3}}Ai''(\zeta)\right]
\]
Remembering the relation (\ref{ker1}) we have for the kernel $W_1(x,x)$ at $x=2+\frac{\zeta}{N^{2/3}}$
\be\label{ker1softedge}
W_1(x,x)= N\left[p^2_{N-1}(x)-p_{N}(x)p_{N-2}(x)\right]\approx 2 \pi\, N^{2/3}\,e^{N}e^{2N^{\frac{1}{3}}\zeta}\rho(\zeta)\,
\ee
where we denoted $\rho(\zeta)=Ai'(\zeta)^2-Ai(\zeta)Ai''(\zeta)$ (which indeed
is proportional to the mean GUE eigenvalue density in the "soft edge" regime).
Then in view of the exact relation $W_2(x,x)=\frac{d}{dx} W_1(x,x)=N^{2/3}\frac{d}{d\zeta}W_1(x,x)$ it follows to the leading order in $N\gg 1$
$W_2(x,x)\approx 2N W_1(x,x) $ for $ x=2+\frac{\zeta}{N^{2/3}}$. This is indeed compatible with the "bulk" relation (\ref{kerker}) $W_2(x,x)\approx Nx W_1(x,x)$ as now $x\approx 2$. We shall see however that for our goals we need to keep the corrections to this approximation:
\be\label{ker2softedge}
W_2(x,x)\approx 2 \pi\,N^{2/3}\,\,e^{N}e^{2N^{\frac{1}{3}}\zeta}\,\left[2N\rho(\zeta)
+N^{2/3}\frac{d}{d\zeta}\rho(\zeta)\right]
\ee
so that
\be\label{softedgeratio}
\frac{W_2(x,x)}{2W_1(x,x)}\approx N+N^{2/3}\frac{1}{2\rho(\zeta)}\frac{d}{d\zeta}\rho(\zeta), \quad x=2+\frac{\zeta}{N^{2/3}}
\ee

In what follows we will also need a similar recursion for $I_{N}(x)$, which is simply $I_{N-1}(x)=-\frac{i}{N}\frac{d}{dx}I_N(x)$. It can be easily derived from the integral representation (\ref{I}) by using the identity $\frac{d}{dx}\left(e^{-\frac{N}{2}(q-ix)^2}\right)\equiv -i\frac{d}{dq}\left(e^{-\frac{N}{2}(q-ix)^2}\right)$ and integrating by parts. This implies

\[
   I_{N}\left(2+\frac{\zeta}{N^{2/3}}\right)\approx \tilde{A}_N\,e^{N^{\frac{1}{3}}\zeta}a(\zeta),\quad \tilde{A}_N=\frac{i^N}{N^{1/3}}\,e^{\frac{1}{2}N}
\]
\be\label{Iassycorrectsoft}
   I_{N-1}\left(2+\frac{\zeta}{N^{2/3}}\right)\approx -i \tilde{A}_N\,e^{N^{\frac{1}{3}}\zeta}\,\left[ a(\zeta)+\frac{1}{N^{1/3}}a'(\zeta)\right],
\ee
\[
 I_{N-2}\left(2+\frac{\zeta}{N^{2/3}}\right)\approx (-i)^2 \tilde{A}_N\,e^{N^{\frac{1}{3}}\zeta}\,\left[ a(\zeta)+\frac{2}{N^{1/3}}a'(\zeta)+\frac{1}{N^{2/3}}a''(\zeta)\right]
\]

Now, we use (\ref{Cauchytr1}) for $x=2+\frac{\zeta}{N^{2/3}}$ and $\epsilon=\frac{\omega \tilde{y}}{N^{2/3}}$:
\[
f_{N+n}\left(2+\frac{\zeta+i\omega \tilde{y}}{N^{2/3}}\right)=-(-i)^{N+n}s_{\omega}^{N+n-1}\sqrt{\frac{N}{2\pi}}
\]
\be\label{Cauchytrsoft}
\times\,\, e^{-\frac{N}{2}\left(2s_{\omega}+\frac{\zeta s_{\omega}+i|\omega|\tilde{y}}{N^{2/3}}\right)^2} I_{N+n}\left(2s_{\omega}+\frac{\zeta s_{\omega}+i|\omega|\tilde{y}}{N^{2/3}}\right)
\ee

Using (\ref{Iassysoft})  we get the leading order expressions for $\omega>0$:
\be\label{fassysoft+}
f_{N}\left(2+\frac{\zeta+i\omega \tilde{y}}{N^{2/3}}\right)\approx B_N\,e^{-N^{\frac{1}{3}}(\zeta+i\omega\tilde{y})}\, a(\zeta+i\omega\tilde{y}), \quad B_N=-\frac{N^{1/6}}{\sqrt{2\pi}}\,e^{-\frac{3}{2}N}\,\quad \omega>0
\ee
where
\be\label{omega+}
a(\zeta+i\omega\tilde{y})=\int_0^{\infty}\,d\tau\,e^{i\tau\zeta+i\frac{\tau^3}{3}}e^{-\omega\tilde{y}\tau}+i
\int_0^{\infty}\,d\tau\,e^{\tau\zeta-\frac{\tau^3}{3}}e^{i\omega\tilde{y}\tau}
\ee

 Similar calculations can be performed for $\omega<0$ (this requires to know asymptotics of $I_N(x)$ around $x=-2$) with the result

\be\label{fassysoft-}
f_{N}\left(2+\frac{\zeta+i\omega \tilde{y}}{N^{2/3}}\right)\approx - B_N\,e^{-N^{\frac{1}{3}}(\zeta+i\omega\tilde{y})}\, \tilde{a}(\zeta+i\omega\tilde{y}), \quad B_N=-\frac{N^{1/6}}{\sqrt{2\pi}}\,e^{-\frac{3}{2}N}\,\quad \omega>0
\ee
where
\be\label{omega}
\tilde{a}(\zeta+i\omega\tilde{y})=\int_0^{\infty}\,d\tau\,e^{-i\tau\zeta-i\frac{\tau^3}{3}}e^{\omega\tilde{y}\tau}-i
\int_0^{\infty}\,d\tau\,e^{\tau\zeta-\frac{\tau^3}{3}}e^{i\omega\tilde{y}\tau}
\ee

Combining these two formulas for any real $\omega$ we can write
\be\label{fassysoft--}
f_{N}\left(2+\frac{\zeta+i\omega \tilde{y}}{N^{2/3}}\right)\approx s_{\omega}B_N\,e^{-N^{\frac{1}{3}}(\zeta+i\omega\tilde{y})}\, \alpha(\zeta,\omega), \quad B_N=-\frac{N^{1/6}}{\sqrt{2\pi}}\,e^{-\frac{3}{2}N}\,
\ee
where
\be\label{anyomega}
\alpha(\zeta,\omega)=\int_0^{\infty}\,d\tau\,e^{is_{\omega}\left(\tau\zeta+\frac{\tau^3}{3}\right)}e^{-|\omega|\tilde{y}\tau}+is_{\omega}
\int_0^{\infty}\,d\tau\,e^{\tau\zeta-\frac{\tau^3}{3}}e^{i\omega\tilde{y}\tau}
\ee

\be\label{fassysoft1}
f_{N-1}\left(2+\frac{\zeta+i\omega \tilde{y}}{N^{2/3}}\right)\approx s_{\omega}B_N\,e^{-N^{\frac{1}{3}}(\zeta+i\omega\tilde{y})}\, \left[\alpha(\zeta,\omega)
+\frac{1}{N^{1/3}}\alpha'(\zeta,\omega)\right],
\ee
where the dash stands for the derivative over $\xi$.

Now we substitute (\ref{fassysoft1}) and (\ref{passycorrectsoft}) to the kernel
\be
F_1\left(x,x+\frac{i\omega \tilde{y}}{N^{2,3}}\right)=f_{N}\left(x+\frac{i\omega \tilde{y}}{N^{2/3}}\right)p_{N-1}(x)-f_{N-1}\left(x+\frac{i\omega \tilde{y}}{N^{2/3}}\right)p_{N}(x), \quad x=2+\frac{\zeta}{N^{2/3}}
\ee
and obtain
\[
F_1\left(2+\frac{\zeta}{N^{2/3}},2+\frac{\zeta+i\omega \tilde{y}}{N^{2/3}}\right)\approx
\]
\be\label{F1soft}
\,e^{-N}\,  e^{-i\,N^{1/3}\omega\,\tilde{y}} s_{\omega}
\left[Ai(\zeta)\,\alpha'(\zeta,\omega)-Ai'(\zeta)\,\alpha(\zeta,\omega)\right],
\ee
A similar calculation gives also:
\[
F_2\left(x,x+\frac{i\omega \tilde{y}}{N^{2/3}}\right)\approx N\,e^{-N}\,  e^{-i\,N^{1/3}\omega\,\tilde{y}} s_{\omega}\times
\]
\be\label{F2soft}
\left\{\left[Ai(\zeta)\,\alpha'(\zeta,\omega)-Ai'(\zeta)\,\alpha(\zeta,\omega)\right]+
\frac{1}{N^{1/3}}\left[Ai''(\zeta)\,\alpha(\zeta,\omega)-\alpha'(\zeta,\omega)\,Ai'(\zeta)\,\right]\right\},
\ee
We see that to the leading order $F_2\left(x,x+\frac{i\omega \tilde{y}}{N^{2/3}}\right)\approx NF_1\left(x,x+\frac{i\omega \tilde{y}}{N^{2/3}}\right)$ in the regime $x=2+\frac{\zeta}{N^{2/3}}$, in full agreement
with the $x\to 2$ limit of (\ref{F2as}). We shall see however that for our goal we need the full expression (\ref{F2soft}).


\begin{thebibliography}{99}
\bibitem{TW} C.A. Tracy, H. Widom {\it Comm. Math. Phys.} {\bf 159} 151  (1994)
\bibitem{Soshnikov} A. Soshnikov {\it Comm. Math. Phys.} {\bf 207} 697 (1999); S. Sodin {\it J. Stat. Phys.}
{\bf 136}, 834 (2009)
\bibitem{Johansson} K. Johansson {\it  Comm. Math. Phys.} {\bf 209} 437 (2000); T. Sasamoto, H. Spohn {\it Nucl. Phys. B}
 {\bf 834} 523 (2010); P. Calabrese, P. Le Doussal, A. Rosso {\it  Europh. Lett.} {\bf  90} 20002 (2010) ; V. Dotsenko {\it  Europh. Lett.} {\bf  90} 20002 (2010)
 \bibitem{Vavilov} M.G. Vavilov et al. {\it Phys. Rev. Lett.} {\bf 86}  874 (2001)
 \bibitem{Fridman} M. Fridman et al. e-preprint arXiv:1012.1282
 \bibitem{QChaos} F. Haake, Quantum Signatures of Chaos,( Springer; 3rd ed. edition , 2010)
\bibitem{ZD} K. Zakrzewski, D. Delande {\it Phys. Rev. E} {\bf 47}  1650 (1993)
\bibitem{FvO} F. von Oppen {\it Phys. Rev. Lett.} {\bf 73} 798 (1994);  F. von Oppen {\it Phys.Rev. E} {\bf 51} 2647 (1995)
\bibitem{FS} Y.V. Fyodorov and H.-J. Sommers {\it Z.Phys.B} {\bf 99} 123 (1995); Y.V. Fyodorov and H.-J. Sommers {\it Phys.Rev. E} {\bf 51} R2719
(1995)
\bibitem{EF} G. Ergun and Y.V. Fyodorov {\it Phys.Rev. E} {\bf 68} 046124 (1995)
\bibitem{spherical} J. M. Kosterlitz, D. J. Thouless, and R. C. Jones  {\it Phys. Rev. Lett.} {\bf 36}, 1217 (1976)
\bibitem{NM}  C. Nadal and S. N. Majumdar, {\it J. Stat. Mech.} {\bf 2011} P04001 (2011)
\bibitem{Jiang} T. Jiang {\it Ann. Prob.} {\bf 34} 1497 (2006)
\bibitem{FStr} Y.V. Fyodorov and E. Strahov {\it J. Phys. A} {\bf 36}, 3203 (2003)
 \bibitem{StrF} E. Strahov and Y.V. Fyodorov {\it Commun. Math.Phys.} {\bf 241} 343 (2003)
\bibitem{Deift} P. Deift, Orthogonal Polynomials and Random Matrices: a Riemann-Hilbert approach. (Courant Lecture Notes in Mathematics, AMS, New York, 1998)
\bibitem{mylec} Y.V. Fyodorov  in "Recent Perspectives in Random Matrix Theory and Number Theory" (London Mathematical Society Lecture Note Series 322; ed. by Mezzardi F.; Snaith N. C.) Cambridge Univesity Press (2005)[arXiv:math-ph/0412017]
\bibitem{FNM} Y.V. Fyodorov, C. Nadal and S. Majumdar, in progress.
\bibitem{AF} G. Akemann and Y.V. Fyodorov {\it Nucl. Phys. B} {\bf 664}[PM] 457 (2003)
\bibitem{BHedge} E. Brezin and H. Hikami {\it Phys. Rev. E} {\bf 62}  3558 (2000)
\bibitem{Fedge} P.J. Forrester {\it Nucl.Phys. B} {\bf 402} 709 (1993)
\bibitem{Gaspard} P. Gaspard, S.A. Rice, H. J. Mikeska and K. Nakamura {\it Phys. Rev. A} {\bf 42 }, 4015 (1990)

\end{thebibliography}
\end{document}